%% file: main.tex
\documentclass[letterpaper,twocolumn,10pt]{article}
\usepackage{usenix2022_SOUPS}
\usepackage{hyperref}
\usepackage{cleveref}
\usepackage{tikz}
\usetikzlibrary{patterns}
\usetikzlibrary{arrows.meta}
\usetikzlibrary{calc, math}
\usepackage{booktabs}
\usepackage{multirow}
\usepackage{comment}
\usepackage{subcaption}
\usepackage{eurosym}
\usepackage{array}
\usepackage{tabularx}
\usepackage{amssymb}
\usepackage{nicefrac}
\usepackage{enumitem}
\usepackage{xcolor}
\usepackage{xspace}
\newcommand{\new}[1]{{#1}}
\newcommand{\underconstruction}[1]{{#1}}
\newcommand{\ready}[1]{{#1}}

\newcommand{\facet}{facet\xspace}
\newcommand{\facets}{facets\xspace}

\newcommand{\security}{non-AML security\xspace}
\newcommand{\Security}{Non-AML security\xspace}

\setlist[enumerate,1]{label={\Roman*}.}

\newcommand{\quoteCode}[1]{`#1'}
\newcommand{\quoteSub}[1]{``\emph{#1}''}
\newcommand{\Sub}[1]{\textit{P{#1}}}

\usepackage[draft,inline,nomargin,index]{fixme}
\fxsetup{theme=color,mode=multiuser}
\FXRegisterAuthor{kg}{Kathrin}{\color{purple}Kathrin}
\FXRegisterAuthor{kk}{Katharina}{\color{cyan}Katharina}
\FXRegisterAuthor{lb}{Lukas}{\color{green}Lukas}
\FXRegisterAuthor{mb}{Michael}{\color{red}Michael}

\def\BibTeX{{\rm B\kern-.05em{\sc i\kern-.025em b}\kern-.08em
    T\kern-.1667em\lower.7ex\hbox{E}\kern-.125emX}}

\begin{document}
%don't want date printed
\date{}

\title{\Large \bf \new{Industrial practitioners' mental models of adversarial machine learning}}

\def\plainauthor{Lukas Bieringer, Kathrin Grosse, Michael Backes, Battista Biggio, Katharina Krombholz}

\author{
%{Anonymous Authors$^\dagger$\\}
%$^\dagger$Anonymous Institute or University \\
% $^\dagger$anonymous Mail % $^\ddagger$kathrin.grosse@unica.it
\and \and \and \and \and
{\rm Lukas Bieringer$^*$}\\
QuantPi
\and
{\rm Kathrin Grosse$^*$}\\
University of Cagliari
\and \and \and \and \and
\and
{\rm Michael Backes}\\
CISPA Helmholtz Center \\ for Information Security
\and
{\rm Battista Biggio}\\
University of Cagliari, \\ Pluribus One
\and 
{\rm Katharina Krombholz}\\%\\
CISPA Helmholtz Center \\ for Information Security
%$^\dagger$QuantPi,$^\ddagger$University of Cagliari,$^\S$CISPA Helmholtz Center for Information Security \\
% $^\dagger$lukas.bieringer@quantpi.com $^\ddagger$kathrin.grosse@unica.it
} % end author

\maketitle
\thecopyright

\begingroup\renewcommand\thefootnote{*}
\footnotetext{First two authors contributed equally.}
\endgroup

\begin{abstract}
Although machine learning is widely used in practice, little is known about practitioners' understanding of potential security challenges. In this work, we close this substantial gap and contribute a qualitative study focusing on developers' mental models of the machine learning pipeline and potentially vulnerable components. 
Similar studies have helped in other security fields to discover root causes or improve risk communication.
Our study reveals two \facets of practitioners' mental models of machine learning security.
Firstly, practitioners often confuse machine learning security with threats and defences that are not directly related to machine learning. Secondly, in contrast to most academic research, our participants perceive security of machine learning as not solely related to individual models, but rather in the context of entire workflows that consist of multiple components. 
Jointly with our additional findings, these two facets provide a foundation to substantiate mental models for machine learning security and have implications for the integration of adversarial machine learning into corporate workflows, \new{decreasing practitioners' reported uncertainty}, and appropriate regulatory frameworks for machine learning security.
\end{abstract}

\input{intro.tex} 

\input{relWork.tex}

\input{methodology.tex}

\input{study.tex}

\input{discussion.tex}

\input{limitations.tex}

\input{concl.tex}

\section*{Acknowledgements}
The authors would like to thank Antoine Gautier, Michael Schilling, the anonymous reviewers and the shepherd for the insightful feedback.
This work was supported by the German Federal Ministry of Education and
Research (BMBF) through funding for the Center for IT-Security,
Privacy and Accountability (CISPA) (FKZ: 16KIS0753) and by BMK, BMDW and the Province of Upper Austria within the COMET program managed by FFG in the COMET S3AI module.

\bibliographystyle{plain}
\bibliography{lit.bib}

\normalsize
\appendix
\input{appendix.tex}

%\appendix
\end{document}

%% file: intro.tex
\section{Introduction}
Adversarial machine learning (AML) studies the reliability 
of learning based systems in the context of an adversary~\cite{barreno2006can,biggio2018wild,papernot2018marauder}.
For example, tampering with some features often suffices to
change the classifier's outputs to a class chosen by the adversary~\cite{Dalvi:2004:AC:1014052.1014066,DBLP:conf/pkdd/BiggioCMNSLGR13,DBLP:journals/corr/SzegedyZSBEGF13}. Analogously, slightly altering the training data enables the attacker to decrease performance of the classifier~\cite{rubinstein2009antidote,biggio2011support}. Another change in the training data allows the attacker 
 to enforce a particular output class when a specified stimulus is present~\cite{ji2017backdoor,chen2017targeted}.
Most state-of-the-art attacks and mitigations are
 in an ongoing arms race~\cite{carlini2017adversarial,athalye2018obfuscated,tan2019bypassing}.

\begin{figure*}
\centering
\vspace{-0.2em}
\input{figures/pipeline.tex}
\vspace{-0.5em}
\caption{AML threats within the ML pipeline. Each attack is visualized as an arrow pointing from the step controlled to the point where the attack affects the pipeline.}\label{fig:pipeline}
\vspace{-0.2em}
\end{figure*}
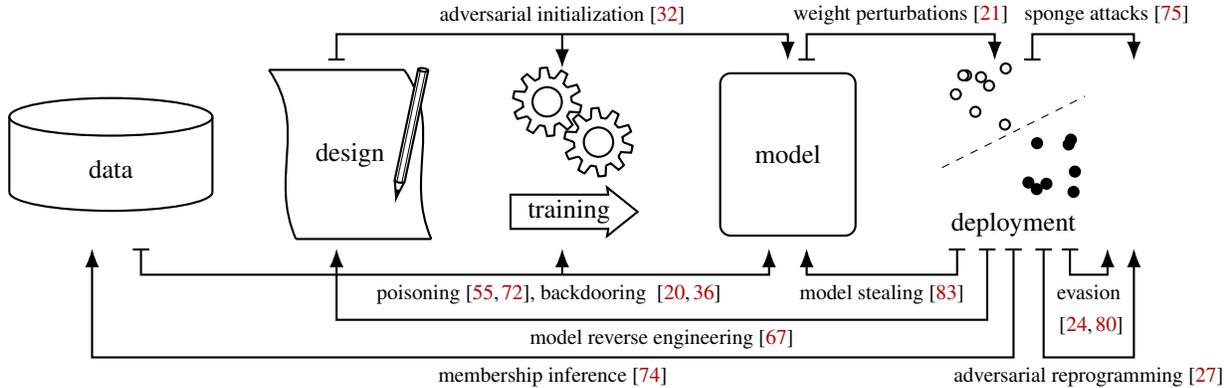

Although machine learning (ML) is increasingly used in industry, \new{very little is known about ML security in practice. At the same time, previous works show that practitioners are  concerned about AML~\cite{mirsky2021threat,kumar2020adversarial}, and failures already occur~\cite{lin2021adversarial},}
very little is known about ML security in practice.
To tackle this question, we conduct a first study to explore mental models of AML.
Mental models are relatively enduring, internal conceptual representations of external systems that originated in cognitive science~\cite{7d020ff455a54632b4f80d0728ac54e6,10.5555/7909}.
 In other security related areas, correct mental models have been found to ease the communication of security warnings~\cite{bravo2010bridging} or enable users to implement security best-practices~\cite{tabassum2019don}. Mental models also serve to 
enable better interactions with a given system~\cite{wash2011influencing}, or to design better user interfaces~\cite{gallagher2017new}.

Our methodology builds upon these previous works by using qualitative methods to investigate the perception of  vulnerabilities in ML applications. \new{More concretely, we conducted 15 semi-structured interviews and  drawing tasks with industrial practitioners from European start-ups and coded both drawings and the transcripts of the interviews.} 
As the first work in this direction, we 
lay the foundations for \new{practitioners'} mental models of AML by describing two \facets of these models.
The first concerns the separation of ML related security (AML) and security unrelated to ML (\security).
In many cases, the borders between these two fields are blurry: a \underconstruction{participant} may start talking about evasion and finish the sentence with a reference to cryptographic keys. 
The second \facet concerns the view of the ML model within a project.
In contrast to the focus on an isolated
model in AML research~\cite{biggio2018wild,barreno2006can,papernot2018marauder,carlini2017adversarial,athalye2018obfuscated},
our practitioners often describe one or more pipelines with potentially several applications of ML.
Finally, we found more \facets which are left for an in-depth investigation by future work. These include the application setting, prior education, and the perceived relevance of AML.

Our interviews showed that most of our \underconstruction{participants} lack an adequate and differentiated understanding to secure ML systems in production. 
At the same time, more than a third of our participants feels insecure about AML. These concerns seem justified as we found evidence for semi-automated fraud on ML systems in the wild. 
However, our findings have more practical implications.
Our results allow us to address the current lack of understanding by (\textbf{I}) \new{increasing awareness for AML and decreasing uncertainty about AML,} (\textbf{II}) developing tools that help practitioners to assess and evaluate security of ML applications, and (\textbf{III}) drafting regulations that contain adequate security assessments and reduce insecurity about AML. 
However, more work is needed to understand the individual and shared mental models of practitioners and assess the real world security risks when applying ML. 

%% file: figures/pipeline.tex
\begin{tikzpicture}
  % definitions  
  \tikzset{myline/.style = {thick, black}}
  \tikzset {
    partial ellipse/.style args={#1:#2:#3}{
      insert path={+ (#1:#3) arc (#1:#2:#3)}
    }
  }

  % data
  \begin{scope}[xshift=-6cm,xscale=0.9]
    \draw [myline] (0, 0.5) ellipse (1.5cm and 0.25cm);
    \draw [myline] (0, -0.5) [partial ellipse=180:360:1.5cm and 0.25cm];
    \draw [myline] (-1.5, -0.5) -- (-1.5, 0.5);
    \draw [myline] (1.5, -0.5) -- (1.5, 0.5);
    \draw node at (0, -0.2) {data};    
  \end{scope}

  % design
  \begin{scope}[xshift=-3cm,yscale=0.75,xscale=0.9]
    \draw [myline] (-1,1.5) -- (1,1.5) edge[out=-60, in=110] (1.4,-1.5);
    \draw [myline] (1.4,-1.5) -- (-0.6,-1.5) edge[out=110, in=-60] (-1,1.5);
    \draw node at (0.2, 0.) {design};
    \begin{scope}[rotate=-10,xshift=1cm,yshift=0.7cm]
      \fill [fill=white] (-0.075,-1) rectangle (0.075,1);
      \draw [myline] (0.075,-1) -- (0.075,1);
      \draw [myline] (-0.075,-1) -- (-0.075,1);
      \draw [myline] (0, 1) ellipse (0.075cm and 0.03cm);
      \draw [myline, fill=white] (-0.075,-1) -- (0,-1.3) -- (0.075,-1);
      \draw [myline] (0, -1) [partial ellipse=180:360:0.075cm and 0.03cm];
      \draw [] (0.0375,-1.025) -- (0.0375,0.99);
      \draw [] (-0.0375,-1.025) -- (-0.0375,0.99);
      \draw [myline, fill=black] (-0.01875,-1.175) -- (0,-1.3) -- (0.01875,-1.175) -- cycle;
    \end{scope}
  \end{scope}

  % training
  \begin{scope}[xshift=-0.2cm,yshift=0.7cm,scale=0.65]
    \draw[myline]
    \foreach \i in {1,2,...,10} {%
      [rotate=(\i-1)*36]  (0:0.5)  arc (0:12:0.5) -- (18:0.7)  arc (18:30:0.7) --  (36:0.5)
    };
    \draw [myline] (0,0) circle (0.3);
    \begin{scope}[xshift=1.05cm,yshift=-0.825cm]
      \draw[myline]
      \foreach \i in {1,2,...,10} {%
        [rotate=(\i-1)*36]  (0:0.5)  arc (0:12:0.5) -- (18:0.7)  arc (18:30:0.7) --  (36:0.5)
      };
      \draw [myline] (0,0) circle (0.3);
    \end{scope}
    \begin{scope}[xshift=0.25cm, yshift=-2.25cm]
      \draw [myline] (-1.,0.35) -- (1,0.35) -- (0.975,0.5) -- (1.6, 0) -- (0.975,-0.5) -- (1,-0.35) -- (-1,-0.35) -- cycle;
      \draw node at (0.2,0.) {training};
    \end{scope}
  \end{scope}

  % model
  \begin{scope}[xshift=3cm,scale=0.9]
    \draw [myline, rounded corners] (-1,-1.2) rectangle (1,1.2);
    \draw node at (0,0) {model};
  \end{scope}

  % application
  \begin{scope}[xshift=6cm,yshift=0.3cm,scale=0.95]
    \foreach \x in {1, ..., 8}{
      \draw [myline,fill=white] (0.4*rand-0.5,0.4*rand+0.5) circle (0.07);
      \draw [myline,fill=black] (0.4*rand+0.5,0.4*rand-0.5) circle (0.07);
    }
    \draw [dashed] (-1,-0.5) -- (1,0.5);
    \draw node at (0,-1.3) {deployment};
  \end{scope}

  \newcommand{\attackarrow}[4]{
    \draw [thick, |-{Latex[length=2.5mm]}] (#1, -1.25) -- (#1, -1.25 - #3) -- node[below,align=center]{#4} (#2, -1.25 - #3) -- (#2, -1.25);
  }
  \newcommand{\attackarrowa}[4]{
    \draw [thick, |-{Latex[length=2.5mm]}] (#1, 1.25) -- (#1, 1.25 + #3) -- node[above]{#4} (#2, 1.25 + #3) -- (#2, 1.25);
  }

  \attackarrow{-5.6}{2.75}{0.35}{\hspace{25mm}{\footnotesize poisoning~\cite{rubinstein2009antidote,liu2017trojaning}, backdooring ~\cite{ji2017backdoor,chen2017targeted}}}
  %\attackarrow{-5.75}{0}{0.5}{}
  \fill [fill=white] (-3.1, -1.8) rectangle (-2.9, -2.2);
  \attackarrow{5.25}{3.25}{0.35}{{\footnotesize model stealing \cite{DBLP:conf/uss/TramerZJRR16}}}
  \attackarrow{5.65}{-3.}{0.95}{{\footnotesize model reverse engineering \cite{joon2018towards}}}
  \attackarrow{6}{-6.25}{1.45}{{\footnotesize membership inference \cite{2016arXiv161005820S}}}
  \attackarrow{6.75}{7.25}{0.35}{{\footnotesize evasion}\\{\footnotesize \cite{Dalvi:2004:AC:1014052.1014066,DBLP:journals/corr/SzegedyZSBEGF13}}}
    \attackarrow{6.4}{7.6}{1.45}{{\footnotesize adversarial reprogramming~\cite{2018arXiv180611146E}}}
  \attackarrowa{-3}{3}{0.35}{{\footnotesize adversarial initialization~\cite{grosse2019adversarial}}}
  \draw [thick, -{Latex[length=2.5mm]}] (0,1.6) -- (0,1.15);
  \draw [thick, -{Latex[length=2.5mm]}] (0,-1.6) -- (0,-1.25);

  \attackarrowa{3.25}{5.75}{0.35}{{\footnotesize weight perturbations~\cite{cheney2017robustness}}}
    \attackarrowa{6.25}{7.6}{0.35}{{\footnotesize \hspace{2em} sponge attacks~\cite{shumailov2020sponge}}}
\end{tikzpicture}

%% file: relWork.tex
%!TEX root = ./main.tex

\section{Background and related work}\label{sec:background}
In this section, we review related work on AML and recall different attacks that have recently been discussed. We also review literature on mental models with regard to human-computer interaction, usable security and ML.

\subsection{Adversarial machine learning}\label{sec:aml}
AML studies the security of ML algorithms~\cite{barreno2006can,biggio2018wild,papernot2018marauder}. 
We attempt to give an informal overview of all attacks in AML, and additionally illustrate them in Figure~\ref{fig:pipeline}. 

\textbf{Poisoning/backdooring.} Early works in poisoning altered the training data~\cite{rubinstein2009antidote} or labels~\cite{biggio2011support} to decrease accuracy of the resulting classifier, for example SVM. For deep learning, due to the flexibility of the models, introducing backdoors is more common~\cite{ji2017backdoor,chen2017targeted}. Backdoors are chosen input patterns that reliably trigger a specified classification output. 
Defending such backdoors has lead to an arms race~\cite{tan2019bypassing}.

\textbf{Evasion/adversarial examples.} 
Early work in evasion decreased the test-time accuracy  of spam classification~~\cite{Dalvi:2004:AC:1014052.1014066}. It was later shown that also more complex  models change their output for small, malicious input perturbations~\cite{DBLP:conf/pkdd/BiggioCMNSLGR13,DBLP:journals/corr/SzegedyZSBEGF13}.
Albeit all classifiers are principally vulnerable towards evasion, recent works focus on the 
 arms race in deep learning~\cite{carlini2017adversarial,athalye2018obfuscated}. 

\textbf{Membership inference.} 
After first inferring attributes~\cite{ateniese2015hacking} of the training data, research later showed that entire points can be leaked from a model~\cite{2016arXiv161005820S}. More concretely, the attacker deduces, given the output  of a trained ML model, whether a data record was part of the training data or not. 
As for other attacks, numerous defenses are being proposed~\cite{nasr2018machine,jia2019memguard}.

\textbf{Model stealing.} Tram{\`{e}}r et al.~\cite{DBLP:conf/uss/TramerZJRR16} recently introduced model stealing. During this attack, the attacker copies the ML model functionality without consent of the model's owner. 
The attacker, given black box access to the original model, tries to reproduce a model with similar performance. 
As for the previous attacks, mitigations have been proposed~\cite{juuti2019prada,orekondyprediction}.

\textbf{Weight perturbations.} 
Fault tolerance of neural networks has long been studied in the ML community~\cite{neti1992maximally,breier2018practical}. Recently, maliciously altered weights are used to
 introduce a specific backdoor~\cite{ji2018model}.
 Few works exist to defend malicious change to the weights in general, not only related to backdoor introduction~\cite{stutz2020mitigating,weng2020towards}.

For the sake of completeness, we conclude with a description of additional, recent attacks,
some of which are part of our questionnaires (see Appendix~\ref{app:quesAfter}). 
In \textbf{adversarial initialization}, the initial weights of a neural network\footnote{Classifiers with convex optimization problems (for example SVM) cannot be targeted, as the mathematical solution to the learning problem does not depend on the initial weights.} are targeted to harm convergence or accuracy during training~\cite{grosse2019adversarial,liu2019bad}. 
 In \textbf{adversarial reprogramming}, an input perturbation mask forces the classifier at test time to perform another classification task than originally intended~\cite{2018arXiv180611146E}.
 For example, a cat/dog classifier is reprogrammed to classify digits.
 In \textbf{model reverse engineering}, crafted inputs allow to deduce from a trained model the usage of dropout and other architectural choices~\cite{joon2018towards}. Finally,
\textbf{sponge attacks} aim to increase energy consumption of the classifier at test time~\cite{shumailov2020sponge}. 
 
 \textbf{Practical Relevance of AML.} In general, AML research has been criticized for the limited practical relevance of its threat models~\cite{gilmer2018motivating,evtimov2020security}.
 A possible reason is our lack
 of knowledge about AI security in practice. Few works attempt to tackle this gap, including
 for example Lin and Biggio~\cite{lin2021adversarial}. They give an overview about AI attacks that were carried out in practice based on AI related incidents covered in newspapers.
 Furthermore, Boenisch et al.~\cite{boenisch2021never} conducted a survey and developed an awareness score, which however encompasses AML, privacy, and \security. 
 Concerning 
 which threats are relevant in practice in industry,
 Kumar et al.~\cite{kumar2020adversarial} and Mirsky et al.~\cite{mirsky2021threat} found that practitioners are most concerned about model theft and poisoning. 
 Yet, in academia, most work focused on evasion so far. 
To shed more light on AML in practice, we interview industrial practitioners and take a first step towards a theory of mental models of AML. To this end, we now introduce and review mental models.

\subsection{Mental models}
Mental models are relatively enduring and accessible, but limited, internal conceptual representations of external systems~\cite{doyle1998mental} that enable people to interact with given systems. Hence, the field of human computer interaction (HCI) studied this concept quite early~\cite{sasse1991t}. Mental models, most recently, saw an increasing relevance in usable security. We now recall prior application scenarios and highlight relevant conceptual contributions in the context of security and ML. 

\textbf{Mental models in HCI and usable security.} 
The relevance of mental models has been subject to a lengthy debate in HCI research~\cite{volkamer2013mental, staggers1993mental}. In many cases, the focus was to capture, depict and analyze mental models of specific objects of investigation. Examples of topics include, but are not limited to, the design of online search applications~\cite{bates1989design}, interface design~\cite{khaslavsky1998integrating}, and interfaces for blind people~\cite{donker2002design}. Research in usable security has recently focused on mental models of security in general~\cite{wash2011influencing, anellend}, privacy in general~\cite{renaud2014doesn}, security warnings~\cite{bravo2010bridging}, incident response~\cite{redmiles2019should}, the internet \cite{kang2015my}, the design of security dashboards~\cite{maier2017influence}, the Tor anonymity network~\cite{gallagher2017new}, privacy and security in smart homes~\cite{tabassum2019don,zeng2017end},
encryption~\cite{wu2018tree,abu2018exploring},
HTTPS~\cite{krombholz2019if}, and cryptocurrency systems~\cite{255652}. 

With regard to the respective object of investigation, these contributions paved the way for improvements of user interface designs~\cite{gallagher2017new}, adequate security communication~\cite{bravo2010bridging}, as well as the development of security policies and implementation of best-practices~\cite{tabassum2019don}. It has been argued that security mental models contain structural and functional properties~\cite{wu2018tree}. For each application, users develop a cognitive representation of its inherent components, their interconnection and correspondingly possible security threats. This representation helps them to understand where threats could emerge and how they could take effect. Mental models evolve dynamically upon individual interaction with a given application~\cite{binns2018s}.

\textbf{Mental models in ML.}
In order to interact with an ML application, humans need a mental model of how it combines evidence for prediction~\cite{nguyen2018believe}. This is all the more important for ML-based applications which often inherit a certain opacity. As Lage et al.~\cite{lage2019human} pointed out, the number of necessary cognitive chunks is the most important type of complexity in order to understand applications. During interaction with black-box processes, humans strive for reduced complexity which may lead to the development of inaccurate or oversimplified mental models~\cite{kaur2019interpreting, hitron2019can}. 

A dedicated line of research therefore elaborates on the relevance and nature of mental models in the context of explainable artificial intelligence. %(XAI). 
Mental models have been found to serve as scaffolds not only for a given ML application~\cite{villareale2021understanding}, but also for its embedding in organizational practices~\cite{zhang2020data}. For data science teams, these workflows usually consist of predefined steps (Figure~\ref{fig:pipeline}) and necessitate interpersonal collaboration~\cite{nahar2021collaboration}. Following Arrieta et al.~\cite{arrieta2020explainable}, we argue that individual collaborators within these teams (e.g.\new{,} ML engineers, software engineers) develop separate internal representations of a given workflow or application. The need for appropriate mental models thereby increases with the enlarged scope of ML applications~\cite{lakkaraju2016interpretable} and involved stakeholders~\cite{suresh2021beyond, Langer_2021}.

Recent work in this line of research called for qualitative studies at the intersection of the HCI and ML communities, to better understand the cognitive expectations practitioners have on ML systems~\cite{kaur2019interpreting, 10.1145/3351095.3375624}. Suchlike studies seem all the more relevant as various industry initiatives propagate a human-centric approach to AI, explicitly referring to mental models.\footnote{e.g.\new{,} https://pair.withgoogle.com/chapter/mental-models/} However, the current scientific discourse lacks a dedicated consideration of cognition in AML. In order to fill this gap, we present the first qualitative study to elicit mental models of adversarial aspects in ML.

%% file: methodology.tex
\section{Methodology}\label{sec::methodology}
This section describes the design of our semi-structured interviews, the drawing task, our recruiting strategy, the participants, and the data analysis. Our methodology was designed to investigate the perception of ML security and is, to the best of our knowledge, the first mental model study of AML.

\subsection{Study design and procedure}
To assess participants' perceptions, we conducted semi-structured interviews enriched with drawing tasks. 
We draw inspiration 
from recent work in usable security which also investigated  mental models~\cite{wu2018tree, krombholz2019if}. 

Before the interview, participants were informed about the general purpose of our study and the applied privacy measures. 
We further assured each participant that their answers would not be judged. 
Participants were then instructed to complete a questionnaire on demographics, organizational background and a self-reflected familiarity with field-related concepts (Appendix~\ref{app:questionnaires}) before the interview. This questionnaire was filled with or without the authors' presence. The answers 
have later been used to put participants' perceptions in context to their organizational and individual background.

The threefold structure of our interviews covered 1) a specification of a given ML project a participant was involved in, 2) the underlying ML pipeline of this project and 3) possible security threats within the project. We chose this approach as the different attack vectors form part of the ML-pipeline as shown in Section~\ref{sec:aml}. 
The detailed interview guideline can be found in Appendix~\ref{app:interview}. As a last step of our interviews, we confronted the participants with exemplary attacker models for some of the threats considered 
relevant in industrial application of ML~\cite{kumar2020adversarial}. To assess practitioners' understandings of these threats, study participants had to elaborate on these attack vectors within their specific setup (Appendix~\ref{app:SelAVs}). 

\input{subject_table}

To assess the participants' knowledge about (A)ML in general,
participants were asked to fill an additional questionnaire after the interview (Appendix~\ref{app:quesAfter}). In this questionnaire, we tested general knowledge in ML and independently asked for a self-reflected familiarity rating with some of the attacks we discussed in Section~\ref{sec:aml}. This questionnaire was handed to the participants after the interview as to avoid priming.

We conducted one pilot interview to evaluate our study design. This first participant met all criteria of our target population in terms of employment, education and prior knowledge.
As his explanations and drawings matched our expectations,
we only added a specific question regarding the collaborators within a given ML-based project. 

The average interview lasted 40 minutes and was jointly 
conducted by the first two authors of this paper \new{between April and July 2019}. To
minimize interviewer biases, we equally distributed the interviews,
where one author was
the lead interviewer and the other took notes.
Due to the COVID-19 pandemic, interviews were conducted remotely and relied on a freely available digital whiteboard\footnote{\url{https://awwapp.com/}}. 

\subsection{Recruitment}
Recruitment for a study on applied ML in corporate environments presents a challenge, as only a small proportion of the overall population works with ML. 
Furthermore, the topic touches compliance and intellectual property of participating organizations. Hence, many companies are skeptical about the exchange with third parties. Consequently, many current contributions with industrial practitioners as study participants are conducted by corporate research groups (e.g.\new{,}~\cite{kumar2020adversarial,holstein2019improving}). 

We tried to initiate interviews with two large multinational companies. 
Unfortunately, both denied our request after internal risk assessments. Therefore, we focused on smaller companies where we could present our research project directly to decision-makers and convince them to participate in our study. We relied on the authors' networks (pilot participant, \Sub{11}) and public databases for start-ups (more details in Appendix~\ref{app:databases}) 
to find potential participants and used direct-messaging on LinkedIn and emails to get in contact. 

Recruitment of study participants happened in parallel to interview conduction. Some participants forwarded our interview request to internal colleagues, so that we talked to multiple employees of some participating companies (see Table~\ref{table:subjects}). We aimed to recruit experienced and knowledgeable participants and hence our requirements were a background in ML or computer science and positions such as data scientists, software engineers, product managers, or tech leads. We did not require any prior knowledge in security. \new{After 8 interviews, no new topics (in our case for example new pipeline elements, whether defenses were mentioned, or how attacks were depicted in drawings) emerged. The research team thus agreed after 15 interviews that saturation was reached~\cite{bowen2008naturalistic}, and we stopped recruiting.}
The participants were randomly assigned an ID (a number between 1 and 20) which was used throughout our analysis. All participants were offered an euro 20 voucher as compensation for their time.

\subsection{Participants}
We summarize demographic information in Table~\ref{table:subjects}. One participant, \Sub{10}, did not hand in the questionnaire and is consequently not included in the following statistics. 
14 participants identified as male, one identified as female\new{, our sample is thus skewed towards males when considering ML practitioners~\cite{kaggle}. As previous work found security perception of women and men to exhibit only some differences~\cite{mcgill2021exploring}, this bias is acceptable for a first exploration but should be studied in depth in future work.}
Our participants had an average age of 34 years (standard deviation (STD) 4.27). 
As intended for a first exploration of practitioners' perception of AML, our sample covered various application domains and organizational roles which we now describe in detail. 
 
\textbf{Education and prior knowledge.} The majority of participants (9 of 14) has a PhD, with all participants holding some academic degree.
\new{While our sample skews towards PhDs compared to the overall population of ML practitioners~\cite{kaggle}, previous work reports no correlation between overall education and security awareness~\cite{boenisch2021never}.}
Most participants (12 of 14) reported that they had attended lectures or seminars on ML. Roughly half (6 of 14) reported to have a similar background in security. To obtain a more objective measure we conducted a test about ML knowledge and asked participants to rate their familiarity with AML attacks (details in Appendix~\ref{sec::sanityCheck}).
While we found that all participants were indeed knowledgeable in ML, we found that few attacks were well known to them. 

\textbf{Employment.} Regarding the size of the companies, four participants worked in companies with less than ten employees, five in companies with less than 50 and the remaining six participants in companies with less than 200 employees.
The companies' application areas were as diverse as healthcare, security, human resources, and others. Most participants were working in their current positions 6 years (STD 4.9). Their roles were diverse:
Most (8 of 15) were in managing positions. Three were software or ML engineers, three more researchers. One of the participants stated to be both a researcher and a founder. One participant did not report his role.

Finally, we asked participants to report which goals were part of their companies' AI/ML checklist. Almost all participants (13 of 14) reported that performance mattered in their company. Half (7 of 14) stated that privacy was important. Slightly less than half (6 of 14) focused on explainability and security. Least participants (4 of 14) listed fairness as a goal in their products. To conclude, when interpreting these numbers, one should keep in mind that not all five goals apply equally to all application domains. Furthermore, our sample is too small to derive per area or per company insights, and we thus leave a detailed analysis for future work.

\subsection{Data analysis}
We adopted an inductive approach, 
where we  followed recent work in social sciences and usable security that constructed theories based on qualitative data~\cite{naiakshina2017developers, krombholz2019if}. To distill observable patterns in interview transcripts and drawings, we applied two rounds of open coding, e.g. we assigned one or several codes to sentences, words, or parts of the drawings. We then performed Strauss and Corbin’s descriptive axial coding to group our data into categories and selective coding to relate these categories to our research questions~\cite{strauss1990basics}. Throughout the  coding  process,  we  used  analytic  memos  to  keep  track of thoughts about emerging themes. The final set of codes for interview transcripts and drawings is listed in Appendix~\ref{app:codes}.

As a first step, the first two authors independently conducted open coding sentence by sentence and sketch by sketch. This allowed for the generation of new codes without predefined hypotheses. Afterwards, the resulting codes were discussed and the research team agreed on adding specific codes for text snippets relating to the confusion of standard security and AML.
As a second step, two coders independently coded the data again. After all iterations of coding, conflicts were resolved and the codebook was adapted accordingly. 

During axial coding, the obtained codes were grouped into categories. The first two authors independently came up with proposed categories which have then been discussed within an in-person meeting. While the grouping was undisputed for some of the categories presented in Appendix~\ref{app:codes} (e.g. AML attacks, pipeline elements), for others the research team decided for (e.g. confusion, relevance) or against (e.g. type of ML model applied) the inclusion of a corresponding category only after detailed discussion. In addition, dedicated codes for the perception of participants (e.g. perceives AML as a feature, not a bug or security issue) were added to the codebook. Once the research team agreed on a final codebook, all transcripts and drawings were coded again using corresponding software.\footnote{Available at https://www.taguette.org/ and https://www.maxqda.com/.} In doing so, we aimed for inferring contextual statements instead of singular entities.

The codes and categories served as a baseline for selective coding. Independently, the researchers came up with observations and proposals for specific mental models. Every proposal included a definition of the observation, related codes, exemplary quotes and drawings. The first two authors then met multiple times to discuss the observations and the corresponding relations of codes and categories. \new{The resulting code tree contains 77 interview codes in 12 groups, 44 for drawings (in 5 groups), as depicted in Appendix~\ref{app:codes}.}

\new{Over all interviews, the coders agreed on 989 codes while disagreeing on 136. }\ready{Analogously, there were 275 codes on drawings in total, with 42 disagreements.}
We further calculated Cohen's kappa~\cite{cohen1960coefficient} to measure the level of agreement among the coders. \underconstruction{For interview transcripts, we reached $\kappa= 0.71$; for \ready{the codes assigned to drawings} $\kappa = 0.85$.} 
These values indicate a good level of coding agreement since both values are greater than 0.61~\cite{landis1977measurement}. Given the semi-technical nature of our codebook, we consider these values as substantial inter-coder agreement. Irrespective of this and in line with best practices in qualitative research, we believe that it is important to elaborate how and why disagreements in coding arose and disclose the insights gained from discussions about them. Each coder brought a unique perspective on the topic that contributed to a more complete picture. Due to the diverse background of our research team in AML, usable security and economic geography, most conflicts arose regarding the relevance of technical and organizational elements of transcripts and drawings. These were resolved during conceptual and on-the-spot discussions within the research team. 

\subsection{Expectations of mental models}\label{sec:expectations} 
Given previous work on mental models and ML,
we designed our study in a way that participants would first visualize their 
pipeline and later add corresponding attacks and defenses. For the pipeline, we expected that participants would name basic steps or components, such as data (collection), training, and testing. In general, we assumed participants’ descriptions would vary in technical detail. Regarding AML, one of our motivations to conduct this study was to learn which knowledge our participants had. As a recent phenomenon, AML might not be known at all in practice, although practitioners might be aware of attacks relevant to their specific application. In particular, we did not expect practitioners to depict attacks using a starting and target point, as done in Figure~\ref{fig:pipeline}.

\subsection{Ethical considerations}
The ethical review board of our university reviewed and approved our study design.
We limited the collection of personal data as much as possible and used ID's for participants throughout the analysis. Since all participants were employed at existing companies and partially shared business-critical information, we aimed to avoid company-specific disclosures in this paper. Finally, we complied with both local privacy regulations and the general data protection regulation (GDPR).

%% file: subject_table.tex
\begin{table}[t]
\caption{\new{Participants with their random IDs.} Capital letters denote that participants work in the same company. \new{We denote the application domain and the working experience (Exp.) in years.} Knowledge in ML, Security and AML is encoded as completed lectures (++), seminar/self-study (+) or none ().}\label{table:subjects}
\centering
\resizebox{\linewidth}{!}{%
\begin{tabular}{lrrrrrrrr} 
      \toprule 
&  \multicolumn{2}{c}{Company} & & \multicolumn{4}{c}{Education} \\ 
\cmidrule(l){2-3}  \cmidrule(l){5-8}  
ID & & Application domain & \new{Exp.} & ML & Sec. & AML & Degree  \\ 
\midrule 
   1 &    & Human resources & \new{7} &++ & + &  & PhD\\ 
   3 &   A & Healthcare & \new{0.4} &   &  &  & PhD\\ 
   4 &   B & Cybsersecurity & \new{8} & ++ & + &  & PhD\\ 
   6 &   C & Business intelligence & \new{15} & ++ & ++ & + & PhD\\ 
   7 &    & Computer vision & \new{12} & ++ &  &  & BSc\\ 
   9 &    & Computer vision & \new{9} &  ++ &  &  & MSc\\ 
  10 &    & Cybersecurity & \multicolumn{5}{c}{no questionnaire handed in} \\
   11 &    & Business intelligence & \new{1} & ++ &  &  & PhD\\ 
   12 &    & Retail and commerce & \new{1.4} &   &  & ++ & PhD\\ 
   14 &    & AI as a service & \new{5} & ++ &  & + & PhD\\ 
   15 &    &  Computer linguistics & \new{5} & + & + &  & MSc\\ 
   16 &   C & Business intelligence & \new{3} & ++ & + & + & PhD\\ 
   18 &   A & Healthcare & \new{1.5} & ++ &  &  & PhD\\ 
   19 &   B & Cybersecurity & \new{15} & ++ & ++ & + & MSc\\ 
   20 &   A & Healthcare & \new{1.2} & ++ &  &  & MSc\\ 
\bottomrule 
 \end{tabular}

}
\end{table}

%% file: study.tex
\section{Empirical results}\label{sec:results} 
In this section, we discuss our findings from the interviews and drawings. Given the unexplored nature of mental models of AML, we focus on two main \facets, and discuss additional findings that require a more in depth analysis (in the sense of future work) at the end of this section.

The first of the two main \facets 
is the (mingled) relationship between ML security (AML) and security unrelated to ML (\security). We found that our participants, while not referring to AML and \security interchangeably, still exhibited an often vague boundary between the two topics.
The second \facet concerns the view on ML as part of a larger workflow or product in industry, as opposed to the focus on an isolated model in academia. 
As a description of a high level workflow
requires a high level perspective, we investigate whether it is equivalent to one, which we find not to be true.
Afterwards, we then discuss potential \facets requiring a more in depth investigation: the application setting, prior knowledge of the participant, and the perceived relevance of AML.
 
\subsection{\Security and AML}\label{sec:secvsAML} 
\begin{figure}
    \centering
    \includegraphics[width=0.46\textwidth]{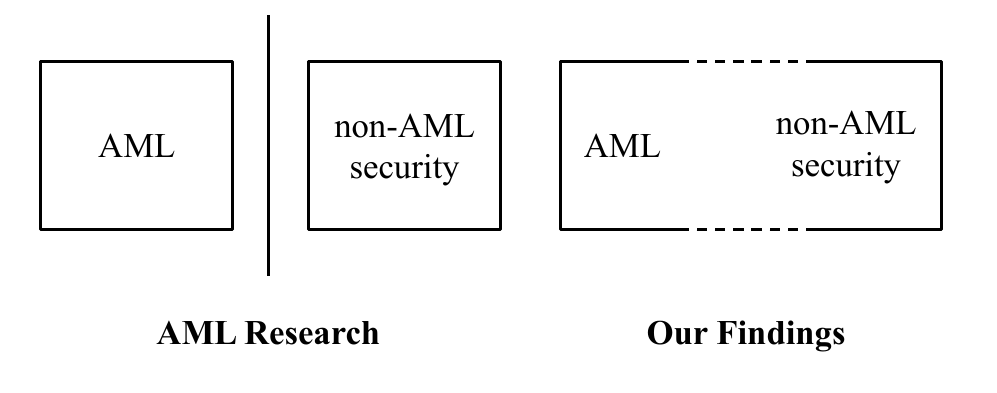}
    \caption{High-level intuition \underconstruction{Section~\ref{sec:secvsAML}}. While in research, \security and AML are rather distinct, our participants do not always clearly distinguish the two fields.}
    \label{fig:Mingling}
\end{figure}

\new{\Security deals with the protection against digital attacks in general. In our case, it encompasses topics like access control, cryptography, malicious code execution, etc.  \Security provides sound solutions by deploying defenses or implementing design choices. In AML, threats are much more connected with the functioning of ML. For many AML attacks, it is unclear which defenses work due to the ongoing arms-race. Although both topics are conceptually  different, we found that }
 our participants did not distinguish between security unrelated to ML and AML, as visualized in Figure~\ref{fig:Mingling}.
In our interviews, on the one hand, the boundary between \security and AML often appeared blurry or unclear, with the corresponding concepts intertwined. 
On the other hand,
there were crucial differences in the perception between \security and AML threats. One difference is that whereas security defenses were often clearly stated as such, AML mitigations\footnote{We are aware that AML is far from being solved, and communicated this to our participants if required. In this study, we define defenses as techniques which increase the difficulty for an attacker, like retraining or explainability.} were often applied without security incentives. Finally, we find a tendency to not believe in AML threats. Many participants denied responsibility, doubted an attacker would benefit, or stated the attack does not exist in the wild. There was no such tendency in \security.

\subsubsection{Mingling AML and \security}
We first provide examples showing that \security and AML were not distinguished by our participants. Afterwards, we investigate if \security and AML are used interchangeably, by investigating the co-occurrence of codes.

\begin{figure*}
    \includegraphics[width=0.98\textwidth]{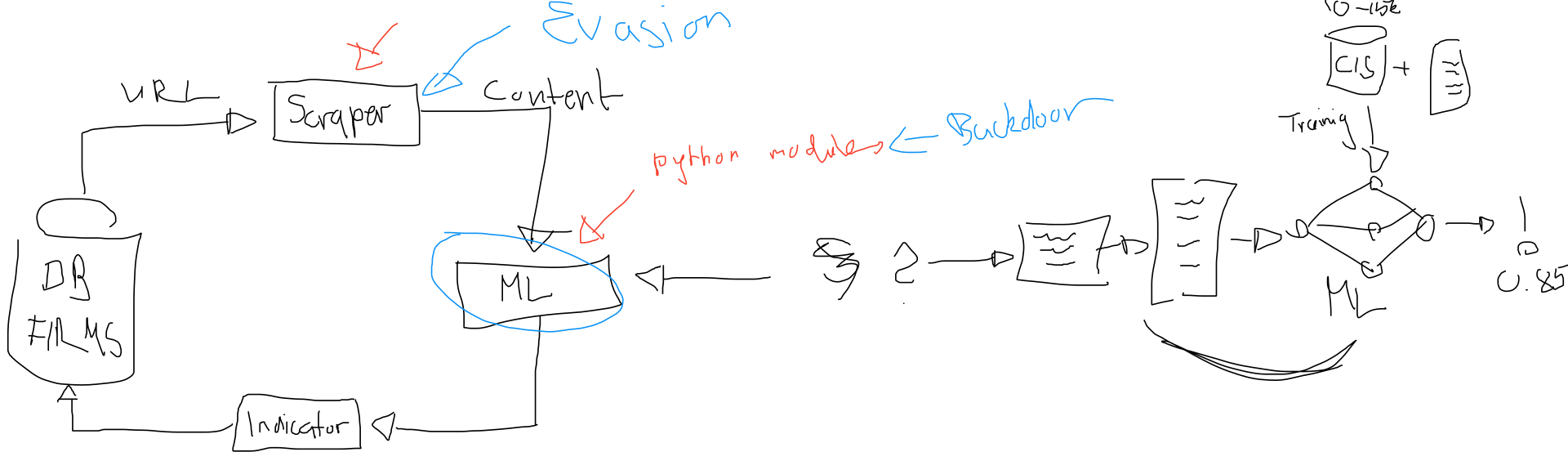}
    \caption{Drawing of \Sub{11}. Red markings were added by the participant before, blue after being confronted with selected attacks.}\label{fig:SecLEvel11}
\end{figure*}

\textbf{Vagueness of the boundary between security and AML.} There are many examples for a vague boundary between \security and AML. For example \Sub{20} reasoned about evasion: \quoteSub{this would require someone to exactly know how we deploy, right? and, where we deploy to, and which keys we use\new{.}} At the beginning, the scenario seems unclear, but the reference to (cryptographic) keys or access tokens shows that the participant has moved to classical security.
Analogously, when \Sub{18} reasoned about membership inference: \quoteSub{but that could be only if you break in [...] if you login in to our computer and then do some data manipulation\new{.}} Again, this participant was reasoning about failed access control as opposed to an AML attack via an API. 
Sometimes, ambiguity in naming confused our participants. For example, \Sub{11} thought aloud: \quoteSub{poisoning [...] the only way to install a backdoor into our models would be that we use python modules that are somewhat wicked or have a backdoor\new{.}} In this case, the term `backdoor' in our questionnaire caused a \security mindset involving libraries in contrary to our original intention to query participants about neural network backdoors. The same reasoning can also be seen in \Sub{11}'s drawing (compare Figure~\ref{fig:SecLEvel11}), where  `backdoor' points to python modules. 
Finally, \Sub{12} stated: \quoteSub{maybe the poisoning will be for the neural network. From our point of view you would have to get through the Google cloud infrastructure\new{.}} From an AML perspective, the attack is carried out via data which is uploaded from the user. Yet, the infrastructure is perceived as an obstacle for the attack. 

\textbf{Correlations between \security and AML attacks.} In the previous paragraph, we showed that the boundaries between AML and \security are blurred in our interviews.
Another example is \Sub{6} reasoning about IP loss: \quoteSub{we are very much concerned I’d say the models themselves and the training data we have that is a concern if people steal that would be bad\new{.}} In this case, it is left out how the attack is performed. Analogously, \Sub{9} remarked: \quoteSub{We could of course deploy our models on the Android phones but we don't want anybody to steal our models\new{.}} 
To investigate whether our participants are more concerned about some property or feature (data, IP, the model functionality) than about how it is stolen or harmed,
 we examined the co-occurrence of AML and \security codes that refer to similar properties in our interviews. For example,  the codes \quoteCode{model stealing} and \quoteCode{code breach} both describe a potential loss of the model (albeit the security version is broader). Both codes
occur together six times, with \quoteCode{code breach} being tagged one additional time. Furthermore, the code \quoteCode{model reverse engineering}, listed only two times, occurs both times with both \quoteCode{model stealing} and \quoteCode{code breach}.
However, not all cases are that clear.
For example \quoteCode{membership inference} and  \quoteCode{data breach} only occur together two times. The individual codes are more frequent, and were mentioned by three (\quoteCode{membership inference}) and eleven (\quoteCode{data breach}) participants.
Analogously, attacks on availability (such as DDoS) in ML and \security were only mentioned once together. 
Such availability attacks were brought up in an ML context twice, in \security four times. 
Codes like \quoteCode{evasion} and \quoteCode{poisoning}, in contrast, are not particularly related to any \security concern. We conclude that AML and security are not interchangeable in our participants' mental models to refer to attacks with a shared goal.

\subsubsection{Differences between AML and \security}\label{sec::AMLvsSec}
In the previous subsection, we found that our participants did not distinguish \security and AML. To show that this is not true in general, we now focus on the differences between the two topics. To this end, we start with the perception of defenses and then consider the overall perception of threats in AML and security. We conclude with a brief remark on the practical relevance of AML.

\textbf{Defenses.} Out of fifteen interviews, in thirteen some kind of defense or mitigation was mentioned; whereas all corresponding interviewees mentioned a \security defense (encryption, passwords, sand-boxing, etc). An AML mitigation appeared in eight. In contrast to security defenses, however, AML defenses were often implemented as part of the pipeline, and not seen in relation to security or AML. As an example, \Sub{9}, \Sub{15}, and \Sub{18} reported to have humans in the loop, however not for defensive purposes. \Sub{10} and \Sub{16} were aware that this makes an attack more difficult. For example, \Sub{16} stated: \quoteSub{maybe this poisoning of the data [...] is potentially more possible.
There, we would have to manually check the data itself. We don’t [...] blindly trust feedback from the user\new{.}} 
Analogous observations hold techniques like explainable models (3 participants apply, 1 on purpose) or retraining (2 apply, additional 2 as mitigation). 
For example, \Sub{14} said: \quoteSub{when we find high entropy in the confidences of the data [...] for those kind of specific ranges we send them back to the data sets to train a second version of the algorithm\new{.}} 
In this case, retraining was used to improve the algorithm, not as a mitigation. 
We conclude that albeit no definite solution to vulnerability exists, many techniques that increase the difficulty for an attacker are implemented by our participants. At the same time, many practitioners are unaware which techniques potentially make an attack harder.

\textbf{Perception of threats.} There is also a huge difference in the perception of threats in \security and AML.
In security, threats were somewhat taken for granted. For example, \Sub{9} was concerned about security of the server's passwords \quoteSub{because anybody can reverse-engineer or sniff it or something\new{.}} Analogously, \Sub{6} said to pay attention to \quoteSub{the infrastructure so that means that the network the machines but also the application layer we need to look at libraries\new{.}}
On the other hand,  
almost a third of our participants (4 of 15) externalized responsibility for AML threats. For example, \Sub{3} said their \quoteSub{main vulnerability from that perspective would probably be more the client would be compromised\new{.}} 
Analogously, \Sub{1} remarked that ML security was a \quoteSub{concern of the other teams\new{.}} In both cases, the participants referred to another entity, and reasoned that they were not in charge to alleviate risks. 
Other reasons not to act include participants not having encountered an AML threat yet, and concluded AML was not relevant. More concretely, \Sub{9} remarked: \quoteSub{we also have a community feature where people can upload images. And there could be some issues where people could try to upload not safe or try to get around something. But we have not observed that much yet. So it's not really a concern, poisoning\new{.}} 
Roughly half of the participants (7 of 15) reported to doubt the attackers' motivation or capabilities in the real world. For example, \Sub{1} said: \quoteSub{I have a hard time imagining right now in our use-cases what an attacker might gain from deploying such attacks\new{.}}
\Sub{20}, who worked in the medical domain, stated: \quoteSub{I’m left thinking, like, why, what could you, achieve from that, by fooling our model. I’m not sure what the benefit is for whoever is trying to do that\new{.}}
Finally, many participants (9 of 15) believed that they have techniques in place which function as defenses. As an in-depth evaluation of which mitigations are effective in which setting is beyond the scope of this paper, we leave it for future work.

\textbf{Practical relevance of AML.} The fact that most participants did not consider AML threats relevant might be an expression of these threats being academic and not occurring in practice. Yet,
 our interviews showed that there are already variants of AML attacks in the wild. More concretely, \Sub{10} stated: \quoteSub{What we found is [...] common criminals doing semi-automated fraud using gaps in the AI or the processes, but they probably don’t know what AML, like adversarial machine learning is and that they are doing that. So we have seen plenty of cases are intentional circumventions, we haven’t quite seen like systematic scientific approaches to crime\new{.}}
 Our participants lack of concern might then be an indicator that harmful AML attacks are (still) rare in practice.

\subsubsection{Summary}
We found that \security and AML were mingled in our participants' mental models:
the boundaries between the corresponding threats were often unclear. Yet, security and AML were not interchangeably used to refer to attacks with a shared goal.
Furthermore, \security threats were treated differently than AML threats: the latter were often considered less relevant. 
Whereas it remains an open question whether AML and \security \emph{should} be treated differently in practice, the fact that they are currently poorly distinguished might due to low  exposure to AML. At the same time, our interviews provided evidence for AML attacks in practice.

\subsection{ML models and ML workflows} \label{sec:techdepth}
\begin{figure}
    \centering
    \includegraphics[width=0.46\textwidth]{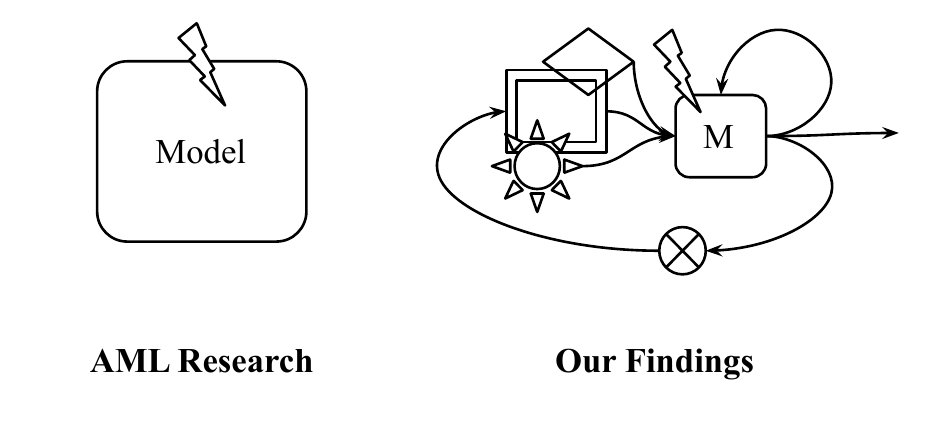}
    \caption{High-level intuition of Section~\ref{sec:techdepth}. While AML research studies individual models, our participants often describe workflows with potentially several models, sometimes even the embedding system of the ML project.}
    \label{fig:ModelViewfigure}
\end{figure}

Many of our participants did not only refer to an ML model, but discussed a workflow or an entire system.
This is in stark contrast to AML research, where models are often studied in isolation, possibly due to a lack of available data. This finding is visualized in Figure~\ref{fig:ModelViewfigure}. In this subsection, we first discuss our participants view on ML models and the described systems. We then investigate whether such views are equivalent to a high level view on ML related projects, and conclude the section with a short discussion on some of our participants' struggles to assess threats at a high level. 

\vspace{-1em}

\subsubsection{Model versus system view} We first focus on the description of the ML model itself. Afterwards, we describe practitioners' views of ML models within larger systems and conclude the section with relating both findings to the technical level of abstraction.

\textbf{ML model perspective.}
The general perception of the ML pipeline (Figure~\ref{fig:pipeline}) seems to affect mainly the relevance of ML-models as such within the pipeline. 
More concretely, 
participants talked about models as pipeline components. 
Many (11 of 15) of our participants presented their projects 
in chronological order or with an implicit flow. Examples are visible in Figure~\ref{fig:SecLEvel11} or Figure~\ref{fig:SecLEvel18}. Moreover, 6 out of 15 participants explained a pipeline not only as being composed by several steps, but remarked potentially several applications of ML within, or that several (different) pipelines exist. 
For example, \Sub{14} reported that \quoteSub{the models are chained one after the other,} and \Sub{7} stated that \quoteSub{we have both like unsupervised training and unsupervised training\new{.}} 
We conclude that often there is not a single model deployed, but data may be processed by several models, potentially in sequential order.

\begin{figure}
    \includegraphics[width=0.98\columnwidth]{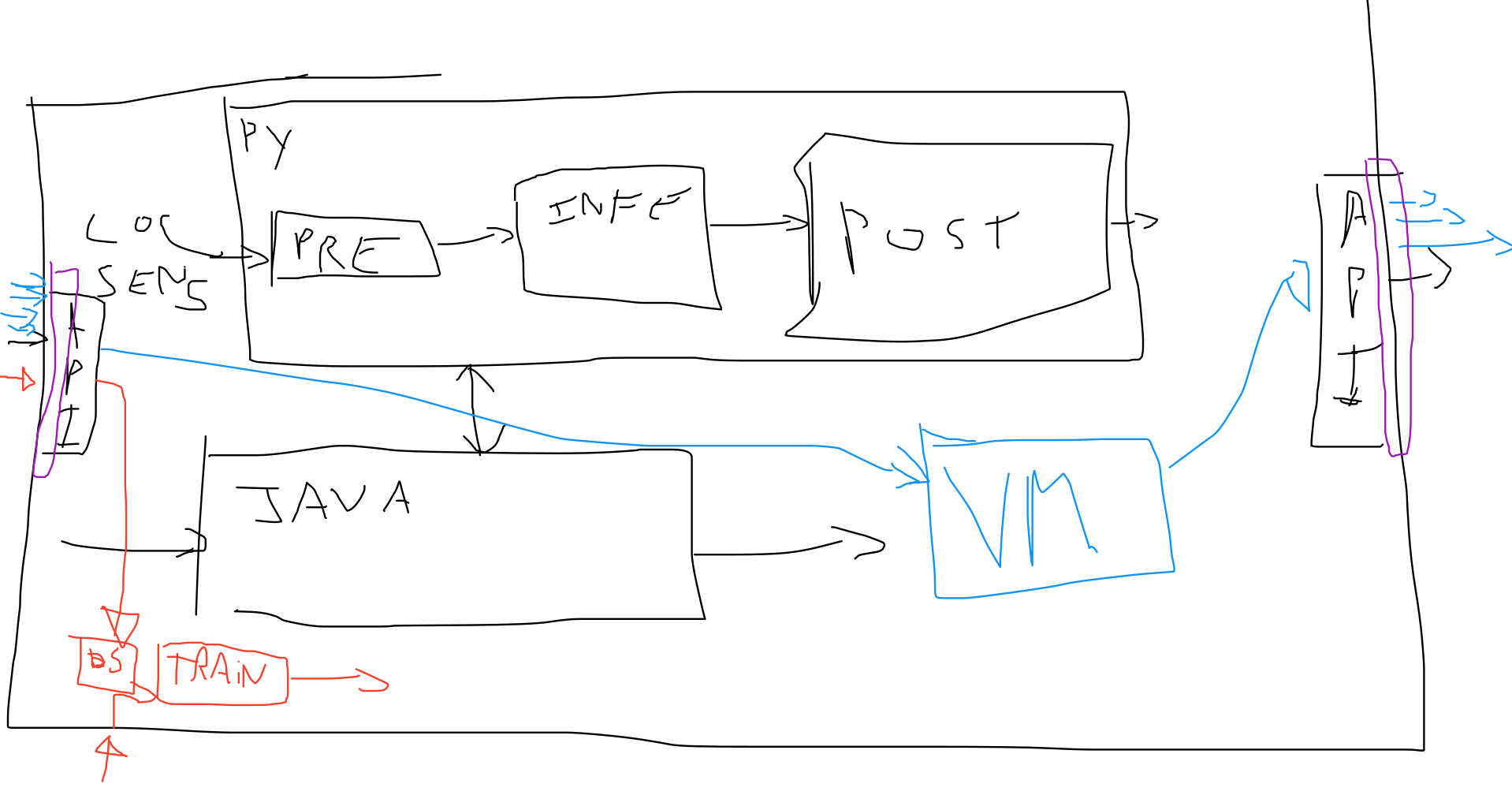}
    \caption{Drawing of \Sub{16}. Colors were added after selected attack were presented to the participant. Red refers to evasion, purple to reverse engineering, blue to membership inference.}\label{fig:SpecAttacks16}
\end{figure}

\textbf{System perspective.}
Moreover, participants showed a strong focus on the surrounding or embedding of their ML-based project. In other words, not only the pipeline around the model was important, but also the surrounding infrastructure of the project.
Out of 15 participants, 5 described their ML pipeline as a classifier as embedded into the larger project context (for example visible in Figure~\ref{fig:SecLEvel11} or Figure~\ref{fig:SpecAttacks16}).  Related to this embedding, in two of the interviews, the topic of technical debt (or long-term maintenance) arose. In this context, \Sub{6} stated: \quoteSub{how [...] we can also have to something that is maintainable in the long term\new{.}}

\subsubsection{Technical abstraction level}
The previous findings 
suggest a high level of technical abstraction in our interviews. While this is true on average, 
some (5 of 15) participants described their project minutely. For example \Sub{12} described their application almost at the code level: \quoteSub{[...] we want to have for each node, that is basically the union of those two columns [...].} However, whereas the same participants also described their project as a workflow, they did not talk about the embedding of the project.
On the other hand, \Sub{18} remarked on their \quoteSub{supervising} (e.g., high level) perspective, yet provided no context. We conclude that our sample does not allow conclusions about the level of technical abstraction and perspective on ML model, which is thus left for future work. 
We did find, however, that a high level perspective seemed to make threat assessment harder for at least 
 some participants. 
Asked to specify a certain threat model, \Sub{19} stated for example: \quoteSub{It's like everywhere. Internal threats, external threats. Trying to mess with the communication, trying to mess if we model something.} 
In a similar manner, \Sub{14} explained that an adversary could \quoteSub{try to put some pythons in non conforming ways to trigger networks.} 
Both descriptions are hard to interpret in technical terms, although both participants seemed aware of security threats in general.
The same problem persists for defenses that our participants apply to encounter AML-specific security threats. \Sub{18}, for example, first explained that \quoteSub{the countermeasures are all in the API.} After rechecking the documentation, the participant was able to provide further details on the applied defenses.

\subsubsection{Summary}
Our findings illustrate an important point which at the same time is very intuitive. Whereas most research papers focus on a single model when investigating ML security, in practice, models are trained and deployed in the context of other models or as components of larger workflows. At the same time, one pipeline may also contain several applications of ML. These views are not to be confused with the technical detail of a projects' description. 
We furthermore find evidence that the right level of detail is crucial to providing useful information.

\subsection{Additional \facets of mental models}\label{sec:cornercases}
Eliciting mental models with only fifteen interviews seems ambitious, in particular in the context of a technique so versatile as ML. In the following, we thus discuss potential aspects of mental models that have to be studied in more depth in future work. These aspects include, but are not limited to the application setting, the effect of prior knowledge, and the perceived relevance of AML. We also found evidence of structural and functional components in our participant’s mental models. As the occurrence of these in AML mental models can be anticipated from prior work in mental models~\cite{wu2018tree}, we leave the corresponding discussion to Appendix~\ref{sec:strucfunc}.

\begin{figure}
    \includegraphics[width=0.98\columnwidth]{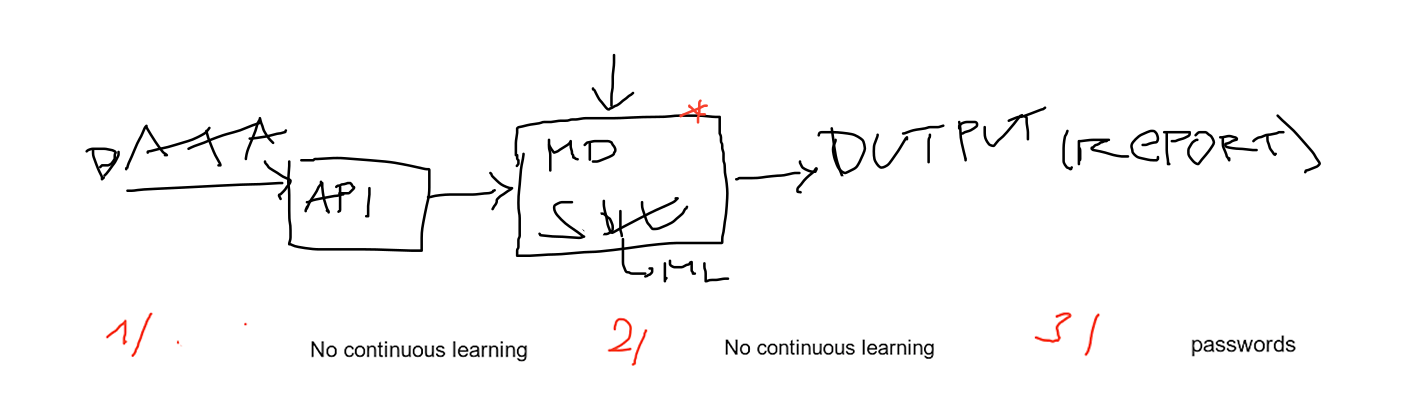}
    \caption{Drawing of \Sub{18}. Red star indicates the most important component of the pipeline, not an attack.}\label{fig:SecLEvel18}
\end{figure}

\subsubsection{Application setting}\label{sec:application}
Our sample is too small to make general statements about the application area. However, since almost a third (4 of 15) participants work in cybersecurity, we attempt to investigate whether working in security affects sensitivity to AML. Hence, we first  divide the participants into security and non-security groups, starting with participants working in security-related fields. 
\Sub{10}, who worked in a setting with cybersecurity reported: \quoteSub{there is some standard AML attacks on ML you can use, but we design our system knowing that very well; on the other hand, we know that there is no perfect security, so, again defense is in monitoring and vigilance, but it’s not something that can be fully automated in our opinion.} \Sub{10} was in general very sensitive towards AML. 
\Sub{4}, also from a cybersecurity setting, was less concerned about evasion: \quoteSub{I can’t imagine yet how it can be applied for real life, for example [...] since we are pretty close on our development.} 
Yet, \Sub{4} also stated the need to gather more information about AML.
Hence, also participants who worked in security-related areas had diverse mental models with respect to concrete attacks. 

Participants from non-security fields have similarly diverse mental models. This diversity is also reflected in the drawings. \Sub{11} (Figure~\ref{fig:SecLEvel11}) added some attacks (in red) before we provided explanations of evasion, backdooring and membership inference (added in blue). \Sub{18} (Figure~\ref{fig:SecLEvel18}), on the other hand, did not add any threats in their drawing. Analogously, opinions also differ in the interviews; e.g.,
\Sub{15} who worked in an non-security setting, was aware of security issues: \quoteSub{one interesting thing of course is that the solution is in some ways constraint by adversarial security considerations so for example you cannot use natural language generation very much because of potential adversarial behavior.} 
On the other hand, and confirming the drawing, \Sub{18} reported that \quoteSub{we do not really protect the machine learning part.} 
Investigating the diversity of mental models induced by the application area in more depth is thus left for future work.

\subsubsection{Prior knowledge}\label{sec:priorknowledge}
Another potential factor on a practitioner's mental model is knowledge about or exposure to the topic at hand. However, we 
 find no strong relation between education and capability or knowledge about AML in our sample. For example, one participant self-reported high knowledge in AML, but also stated:
\quoteSub{maybe the poisoning will be for the neural network.}  
Here, a general attack, poisoning, is related to a specific model (neural networks). 
On the other end of the spectrum, \Sub{9} did not self-report any knowledge about security or AML, but correctly remarked: \quoteSub{Somebody could send us 100.000 images and collect all the results and try to build a model from that.} We conclude that in our sample, self-reported prior knowledge is not related to AML knowledge.
 Yet, more work is needed to understand more in depth the complex relationship between exposure, education, and mental models of AML.

\subsubsection{Perceived relevance of AML}\label{sec:relevanceAML} 
Last but not least, we found little awareness of AML in our sample. As already discussed in section~\ref{sec::AMLvsSec}, this might be a consequence of little exposure to AML attacks in the wild. On the other hand, we found all levels of concern about AML in our sample.
 More concretely, a third of our the practitioners (5 of 15) did not mention AML at all before we explicitly asked. Another third reported that they were not very concerned about AML. For example, \Sub{1} stated that evasion, or \quoteSub{injecting malicious data to basically make the model [...] predict the wrong things} was \quoteSub{a concern that is not as high on my priority list.}
\Sub{15}, analogously, said: \quoteSub{mainly the machine learning pipeline this is the less critical security problem,} reasoning that \quoteSub{simply a performance would be unexpected.} Yet, over a third (6 of 15) of the participants reported to feel insecure about AML when confronted with the topic. Of these six participants, two previously showed low priority on AML, and three did not mention AML at all. 
An example of insecurity is \Sub{4}, who stated they needed \quoteSub{some more research on it.} Some participants, like \Sub{19}, were concerned about specific attacks: \quoteSub{I maybe need to learn more about this membership.} In summary, some practitioners consider AML threats important, whereas some participants did not know AML well, and yet others did not consider it an important threat. From each of these three groups, there was at least one participant that felt not well informed. After the interviews (e.g., off the record) some participants stated that their awareness for AML had increased due to the interview. Many also inquired about defenses against specific threats, further confirming that they were indeed concerned about specific attacks.

%% file: discussion.tex
\section{Future \underconstruction{w}ork}\label{sec:TheoConc}
Our findings expose the lack of knowledge about AML in practice, and thus show the need for additional research at the intersection of AML and cognitive science. In this section, we summarize these potential directions of future work. We first discuss theoretical research on mental models of AML and secondly more practical research that applies findings derived from mental models to AML.

\subsection{A theory on mental models of AML}
Our work is a first step to describe mental models of AML. For well-grounded mental models, more research is needed to investigate different aspects, as discussed in the previous section about the technical detail,  application area and prior knowledge, for example. However, more research is also required concerning the development of mental models, and how a user based threat taxonomy (as opposed to a research based taxonomy) could look like.

\textbf{Temporal evolvement of mental models of AML.} A better understanding about the development of individual mental models could help to assess necessary steps to make practitioners take into account AML. In addition, research on how mental models are shared between various AI practitioners might help to implement adequate defenses within and across corporate workflows. Corresponding starting points can be found in cognitive science~\cite{mohammed2010metaphor}, where the convergence of mental models has been studied as a three-phase process of orientation, differentiation and integration~\cite{kennedy2010merging}. 

\textbf{Inherent threat taxonomies of mental models.} 
Whereas academia has proposed clear threat models in ML security, it is unclear whether or to which degree these are also used or useful in practice. In this context,
 it could be interesting to consider existing taxonomies by Biggio et al.~\cite{biggio2014pattern} and Barreno et al.~\cite{barreno2006can}. These frameworks seem promising to investigate which specific structural elements practitioners consider relevant for specific attack vectors and how they perceive the causal evolution of these attacks. In line with recent work by Wang et al.~\cite{10.1145/3290605.3300831}, such user-centric attack taxonomies might help to understand practitioners' reasoning on AML.

\subsection{Applying mental models to AML}
Secondly, but not less important, is the question how AML research can benefit from the study of mental models and which problems could be tackled in this context. Examples include the usability of AML tools and libraries, a more realistic threat modelling in AML research as well as a general assessment of AML attacks in the wild.

\textbf{Utility and usability of AML tools and libraries.} We found that practitioners’ mental models depend on available and provided information. 
Future research should therefore 
elaborate on the needed specificity of the available  information. 
Furthermore, an evaluation of the available AML tools and libraries with regards to capabilities and needs of industrial practitioners might ease their usage across application domains. In line with recent work on fairness~\cite{lee2020landscape} and ethics~\cite{chivukula2021surveying}, we consider this crucial for designing usable and accessible tools, corporate guidelines and regulations. 

\textbf{Practical threat modelling for AML research.}
As stated in Section~\ref{sec:background}, AML research has been criticized for the limited practical relevance of its threat models~\cite{gilmer2018motivating,evtimov2020security}.
 Mental models could alleviate this issue in two ways. On the one hand, understanding which threats occur in which applications and how they are perceived  helps to shift research towards designing practical and usable defenses. On the other hand, 
a deeper understanding of why \security and AML are mingled allows us to adapt and improve current threat modelling. To this end, however, it is also important to know which threats need to be studied in the first place.

\textbf{AML in the wild.} Given the previous insight and evidence of semi-automated, ML-related fraud, a more detailed assessment of which attacks are conducted in the wild would be beneficial. Future work could investigate this with a focus on different groups of ML practitioners, including for example ML engineers, auditors, and researchers, or dependant on the application. Furthermore, our work outlines that the model perspective usually taken in AML is of limited use in practice. More work is needed to study AML in the context of entire ML pipelines and end-to-end workflows.

\section{Practical \underconstruction{i}mplications}\label{sec:accConc}
Similar to Kumar et al.~\cite{kumar2020adversarial}, we find that most of our \underconstruction{participant}s lack an adequate and differentiated understanding to secure ML systems in production.
Given that we found only reports of semi-automated fraud in our sample \new{in Section~\ref{sec::AMLvsSec}}, the absence of strong AML in practice might explain this lack of knowledge.
Yet, \new{as discussed in Section~\ref{sec:relevanceAML}, 6 of 15 \underconstruction{participant}s felt insecure about ML security}. We thus now discuss the diverse implications of our study on how to tackle these insecurities and the overall lack of knowledge. We start \new{with the question how to raise awareness for AML. Afterwards, } discuss the implications of our findings for the the embedding of AML in corporate workflows and finish with implications for regulatory frameworks of AML.

\new{\textbf{Raising awareness of and increasing confidence about AML.}
Although we did not ask about privacy specifically, the general data protection regulation was often mentioned by our participants. For example, \Sub{6} stated: \quoteSub{we are also subject to GDPR so we cannot just ignore the security aspects of the process.} Like other participants (\Sub{12}, \Sub{18}), \Sub{6} mentioned GDPR before we had asked about membership inference and thus privacy.
Legislation might thus be a tool to increase awareness of AML.
Independently, a third of our participants felt insecure about AML (Section~\ref{sec:relevanceAML}).
Given that several participants reported used software (\Sub{9}, \Sub{14}, for example \quoteSub{TensorFlow}), infrastructure (\Sub{14}) or service provider (\Sub{3}, \Sub{12}, \Sub{20}, for example \quoteSub{Google}), advertising tools to assess AML risks might be helpful for our participants. In particular as 
AML libraries\footnote{For example the Adversarial Robustness Toolbox, CleverHans, RobustBench,
or the SecML library, just to name a few.}, 
but also overviews like the Adversarial ML Threat Matrix\footnote{https://github.com/mitre/advmlthreatmatrix} already exist.
Our findings on the confusion between AML and \security (Section~\ref{sec::AMLvsSec}) suggest  these tools need to either enforce dedicated audits for both AML and \security or combined countermeasures to address both areas jointly. Another solution to the feeling of insecurity, reported by our participants themselves (Section~\ref{sec:relevanceAML}, \Sub{19}: \quoteSub{I maybe need to learn more about this membership}), could be to provide materials for education.}

\textbf{Embedding AML into corporate workflows.} Whereas academia generally studies AML with the perspective of an individual model, in practice, the entire ML pipeline and broader AI workflow need to be considered. As discussed in Section~\ref{sec:techdepth}, in our interviews, \new{for example \Sub{6} and \Sub{16} (see Figure~\ref{fig:SpecAttacks16}) described the entire workflow of their AI application}, whereas other \underconstruction{participant}s focused on the ML pipelines \new{(for example \Sub{18}, as visible in Figure~\ref{fig:SecLEvel18})}. To successfully integrate AML into corporate workflows, however, more effort is needed. All actors working on an ML product need to be able to identify relevant and possible attacks and implementable defenses. \new{Potential factors to consider here are for example different applications areas, as discussed in Section~\ref{sec:application}. Also the existing knowledge of the target audience should be considered, as the in Section~\ref{sec:priorknowledge} discussed variation of knowledge in our sample shows.}

\textbf{Creating appropriate regulatory and standardization frameworks for AML.} Lastly, our study has implications for regulatory approaches that enable appropriate security assessments. The  differences in application \new{(Section~\ref{sec:application})} and prior knowledge \new{(Section~\ref{sec:priorknowledge})} we found 
imply that regulatory frameworks need to find a way to formally encompass these differences with regards to necessary security measures. The currently proposed ‘Legal Framework for AI’ by the European Commission, for example, differentiates certain types of ML applications of which some are prohibited or classified as high-risk and thus require a certain risk management. Furthermore\new{, as discussed in Section~\ref{sec:techdepth}}, our results indicate that it is essential to communicate such frameworks at the right technical abstraction level to encompass both technical ML practitioners and non-technical stakeholders.
Standardization efforts could incorporate this requirement by providing adequate information at multiple mental abstraction levels~\cite{broniatowski2021psychological}. For example, recently proposed frameworks like the NIST Taxonomy and Terminology of AML\footnote{https://nvlpubs.nist.gov/nistpubs/ir/2019/NIST.IR.8269-draft.pdf} 
 explicitly lists references that might help practitioners develop more complex mental models. As \new{mentioned above, a} similar regulatory approach to privacy, the European general data protection regulation, \new{had served as a scaffold for their privacy perception}.

%% file: limitations.tex
\section{Limitations}\label{sec:limitations}
We followed an inductive approach to investigate mental models through qualitative analysis. Hence, the data collected is self-reported and subjected to a coding process. We continued coding and refining codes 
until a good level of inter-coder agreement was reached. Nonetheless, all our findings are subject to interpretation \new{and do not generalize beyond the sample, both of} which is inherent to qualitative analyses. Finally, due to the COVID-19 pandemic, all interviews were conducted remotely and the interface limitations of the digital whiteboard might have impacted the participants' sketches.

\underconstruction{Given the qualitative approach and reached saturation, the small sample size of 15 is indeed acceptable~\cite{wu2018tree, gallagher2017new}.}
\new{Due to the small sample size, however, several factors cannot be addresses in depth, as discussed in Section~\ref{sec:cornercases}. Examples include, but are not limited to, the application setting and the perceived relevance. Ideally, future work provides a more in depth analysis of these topics in a larger quantitative study.} 

All participants were employed at European organizations with $<$200 employees. This is due to the fact that while several multinational companies stated great interest in our research, they denied participation after internal risk assessments. As mental models of ML systems are always embedded in organizational practices~\cite{zhang2020data}, we strongly encourage future research to assess our findings within larger samples including more variety, for example academics, small and large companies.
\new{Given that previous work found differences in general security behavior depending on gender~\cite{mcgill2021exploring}, and cultural background~\cite{kruger2011assessment}, we also strongly encourage a more in depth analysis of these aspects.}

Furthermore, AML itself is a subject of study of which the perception evolves continuously. With an increasing awareness for security within applied machine learning, the findings presented 
can only be valid temporarily.
Machine learning is applied in a wide range of settings. 
Consequently, not all attacks are relevant within each application domain. 
For example, a healthcare setting is subjected to other threats than a cybersecurity setting. 
For the sake of studying abstract \facets of mental models, we did not consider the application in the present work. Yet, we would like to point out the necessity to study this aspect of AML in general.

%% file: concl.tex
\section{Conclusion}\label{sec:conclusion}
Based on our semi-structured interviews with industrial practitioners, we take a first step towards a theory of mental models of AML.
We described two \facets of practitioners' mental models and sketched more \facets as an anchor for in-depth investigation by future work. These include the technical abstraction level, application setting, prior education, and the perceived relevance of AML. We provided more details on the first \facet, or the blurry relationship between AML and \security. These two topics were often mingled, yet not used interchangeably by our participants. 
The second \facet can be understood as a first step to refined threat models in AML research. As apposed to a single model, our participants instead described workflows and relationships between potentially several ML models in a larger system context.

 A clear understanding of the elicited mental models allows to improve information for practitioners and adjustments of corporate workflows.
 More concretely, our results help to \new{raise awareness for AML, thus making practitioners feel less insecure. We further suggest that both application area and prior knowledge are considered when embedding AML into corporate workflows}. Finally, regulatory frameworks might reduce uncertainty about AML and increase the awareness for possible AML threats.
However, a wide range of subsequent research towards an encompassing theory of mental models in AML is still required. Last but nor least, we are convinced that the AML community will benefit from further practical assessment of attacks \underconstruction{in practice}, as our work already provides evidence of semi-automated fraud in the wild.

%% file: appendix.tex
\section{Details on recruiting}\label{app:databases}
We searched online databases like crunchbase
\footnote{\url{https://www.crunchbase.com/} for European companies operating in AI and having raised more than 1 million dollar funding}, AIhubs\footnote{\url{https://www.appliedai.de/hub/2020-ai-german-startup-landscape}}, and lists with promising AI start-ups (for example the list by Forbes\footnote{\url{https://www.forbes.com/sites/alanohnsman/2021/04/26/ai-50-americas-most-promising-artificial-intelligence-companies/?sh=653894c477cf}})
to find potential participants.

\section{Participants' prior knowledge in (A)ML}\label{sec::sanityCheck}
To measure our participants' knowledge in  ML, we constructed a questionnaire based on ML job interview questions\footnote{For example \url{https://www.springboard.com/blog/machine-learning-interview-questions/}}(Appendix~\ref{app:quesAfter}). Given that participants were not informed they had to take a test, we aimed to select a broad range of topics easy to query with multiple choice answers that were not too hard. The questionnaire had 8 questions, with the participants correctly answering on average 6.64 questions (STD 1.14). Guessing would yield an average of 2.66 correct questions. Thus, while we do not know how reliable our questionnaire estimates ML knowledge, we conclude that  our participants are indeed knowledgeable in ML.
 
\begin{figure}[b]
    \centering
    \includegraphics[width=0.47\textwidth]{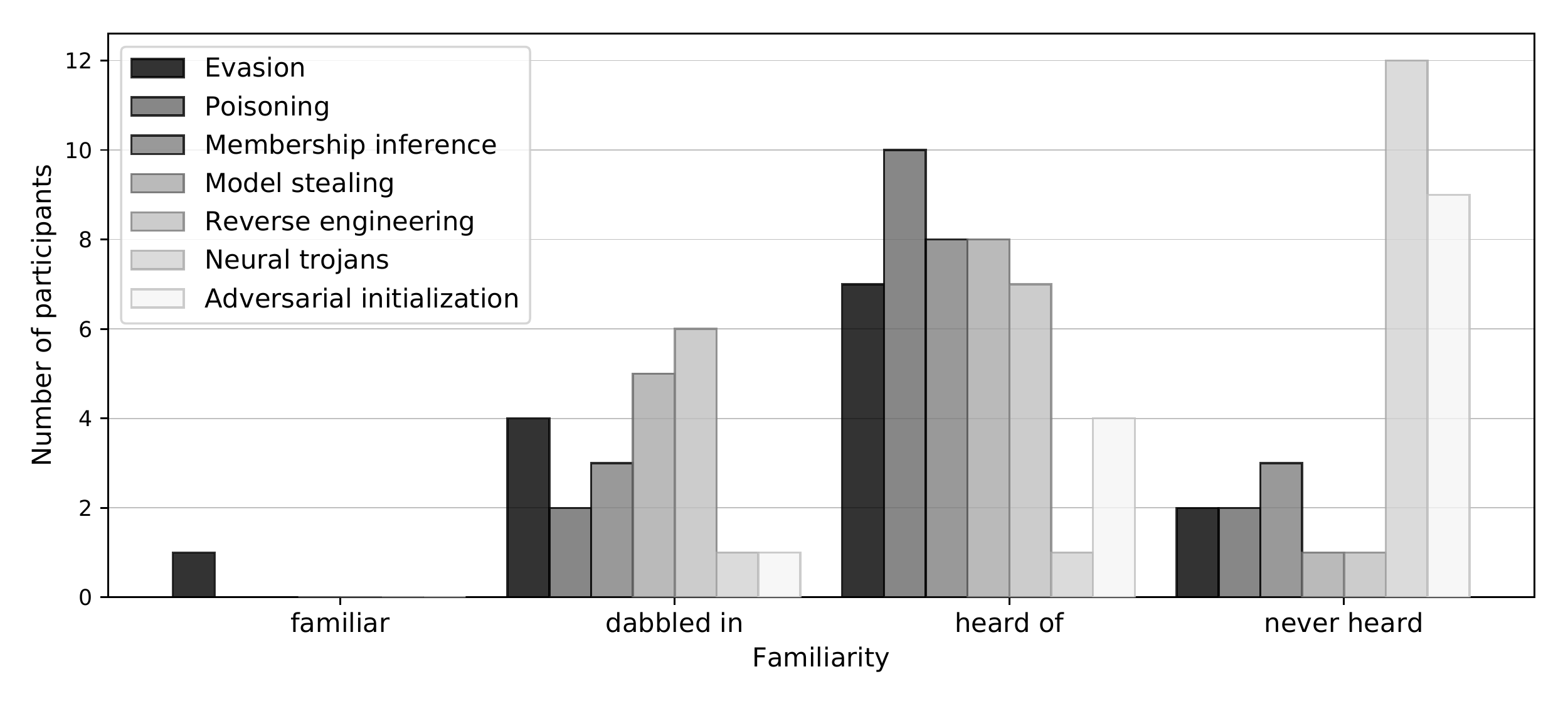}
    \vspace{-6pt}
    \caption{Self-reported familiarity of interviewed participants with different attacks on ML. Total of participants is 14, as one participant did not hand in questionnaire.}
    \label{fig:FamAttacks}
\end{figure}

We also investigated the familiarity of our participants with AML attacks. To avoid priming, we
 asked participants to rate their familiarity after the interview. 
As sanity checks, we added two rather unknown terms, adversarial initialization~\cite{grosse2019adversarial} and neural trojans~\cite{liu2017neural} (similar to backdoors).
 The results are depicted in Figure~\ref{fig:FamAttacks}.
 Only one participant reported to be familiar with one attack (evasion). In general, most participants reported to have heard of most common attacks (evasion, poisoning, membership inference, and model stealing).
As expected for the sanity check, adversarial initialization and neural trojans were largely unknown. 

\section{Interview protocol}\label{app:interview}

Thank you so much for taking the time to give us your perspective on security in machine learning. This study consists in III parts. Part I aims at exploring your role in ML-projects. Part II addresses the underlying machine learning pipeline. In part III, we want to know how you perceive the security of machine learning. In part II and III, please visualize the topics (and relationships between them) that we ask you about. There are no rules, no wrong way to do it, and don’t worry about spelling things perfectly. Nothing is off limits and you can use any feature of the digital whiteboard. After this last part,  we will ask you about your knowledge about security of machine learning before this study.

\vspace{0.5em}

\textbf{Part I: Machine learning project}

$\bullet$ Can you briefly describe what AI- or machine learning-based project you are currently involved in?
	
$\bullet$ Can you tell us a bit more about the goal of this project?

$\bullet$ Who else is involved in this project?

$\bullet$ What is your collaborators role in the project?

\vspace{0.5em}

\textbf{Part II: Machine learning pipeline}

$\bullet$ What kind of pipeline do you currently apply within this machine learning based project?

$\bullet$ Which part of this pipeline is crucial for your business, or identical to your product?

\vspace{0.5em}

\textbf{Part III: Security within project and pipeline}

$\bullet$ Is security something you regularly incorporate into your workflow?

$\bullet$ Have you encountered any issues relating to security in the projects you described?

$\bullet$ Where in the pipeline did these security-related issues originate?

$\bullet$ Can you specify the cause of the security-related issues?

$\bullet$ Can you specify how these security-related issues evolve in your pipeline?

$\bullet$ Which goal pursues an adversary with a such a threat?

$\bullet$ What is the security violation of the threat?

$\bullet$ How specific is the depicted threat?

$\bullet$ Are you aware of any further possible security threats in the scope of your project or pipeline?

$\bullet$ Which countermeasures do you implement against any of the aforementioned threats?

Thank you so much for taking the time to give us your perspective on security in machine learning. 

\vspace{1mm}

\section{Questionnaires}\label{app:questionnaires}
\subsection{Demographics questionnaire}\label{app:demogr}

Thank you for participating in our research study about security in machine learning. Please take a couple of minutes to respond to the following questions. 

\vspace{2mm}

$\bullet$ \makebox[0.3\textwidth]{How old are you? \enspace\hrulefill}
 	  
$\bullet$ What gender do you identify with?

\hspace{1em} $\square$ male  \hspace{3em} $\square$ female \hspace{3em} $\square$ $\rule{2cm}{0.15mm}$  
 
$\bullet$ What is your level of education? (please specify highest)

\hspace{1em} $\square$ Highschool

\hspace{1em} $\square$ \makebox[0.4\textwidth]{Bachelor in \enspace\hrulefill}

\hspace{1em} $\square$ \makebox[0.4\textwidth]{Master / Diploma  in \enspace\hrulefill}

\hspace{1em} $\square$ \makebox[0.4\textwidth]{Training / Apprenticeship in \enspace\hrulefill}

\hspace{1em} $\square$ \makebox[0.4\textwidth]{PhD, area: \enspace\hrulefill}  
 
 $\bullet$ \makebox[0.4\textwidth]{What is your profession? \enspace\hrulefill} 
 
 $\bullet$ \makebox[0.4\textwidth]{What is your role in your team? \enspace\hrulefill} 
 
$\bullet$ How long have you been working in your  

\hspace{1em} \makebox[0.4\textwidth]{ current profession? \enspace\hrulefill} 
 
$\bullet$ What is the number of employees at your 

\hspace{1em} \makebox[0.4\textwidth]{company/organization? \enspace\hrulefill} 
 
$\bullet$ \makebox[0.4\textwidth]{What is the application domain of your product? \enspace\hrulefill} 
 
$\bullet$ Which of these goals are part of your organization's   

\hspace{1em} AI/ML-model checklist?   

\hspace{1em} $\square$ Explainability \hspace{2.5em} 
  $\square$ Fairness \hspace{2.5em}
  $\square$ Privacy  
  
\hspace{1em}  $\square$ Security \hspace{4.84em}
  $\square$ Performance

$\bullet$ In which of these areas have you taken a lecture or intense  

\hspace{1em} course? Please add the title of the course if applicable.

\hspace{1em} $\square$ \makebox[0.35\textwidth]{Machine Learning \enspace\hrulefill}

\hspace{1em} $\square$ \makebox[0.35\textwidth]{Security \enspace\hrulefill}

\hspace{1em} $\square$ \makebox[0.35\textwidth]{Adversarial Machine Learning \enspace\hrulefill}  
 
$\bullet$ In which of these areas have you taken a seminar, or read
 up on? Please add the title of the seminar/book if applicable.

\hspace{1em} $\square$ \makebox[0.4\textwidth]{Machine Learning \enspace\hrulefill}

\hspace{1em} $\square$ \makebox[0.4\textwidth]{Security \enspace\hrulefill}

\hspace{1em} $\square$ \makebox[0.4\textwidth]{Adversarial Machine Learning \enspace\hrulefill}  

\subsection{Attacks used in Interviews}\label{app:SelAVs}

Please read through the following selection of attack vectors and machine learning and explain whether you consider them relevant in your specific project. If yes, please add them to your sketch in a different color.

\textbf{Evasion/ Adversarial Examples.} This attack targets a model during deployment. The goal of the attacker is to fool the model: changing its output significantly by altering the input only slightly. An example is to change a picture containing a dog, present it to a cat-dog-classifier, and the model’s output changes from dog to cat.

\textbf{Poisoning.} This attack targets the training or optimization phase of the model. The goal of the attacker is to either decrease accuracy significantly, or to install a backdoor. An example is a cat-dog classifier that always classifies images containing a smiley as cat.

\textbf{Privacy/ Membership Inference.} This attack targets a model at test-time. The attacker’s goal is to identify individual samples from or even the whole training set. An example is to measure the confidence on an input, as some algorithms tend to be more confident on data they have seen during training. Also over-fitting eases to determine what a classifier was trained on.

\subsection{ML quiz}\label{app:quesAfter}
Please answer the following questions about ML.
 For each question, please tick \textbf{at least} one box.
 
\textbf{Question 1.} Which loss is used to train DNN?

\hspace{1em} $\square$ $0$/$1$-loss.

\hspace{1em} $\square$ Cross-entropy loss.

\hspace{1em} $\square$ Hinge-loss.  

\textbf{Question 2.} What is the difference between classification and regression?

\hspace{1em} $\square$ The kind of labels we fit: reals vs discrete classes.

\hspace{1em} $\square$ Regression is the name of classification in psychology / medical science.

\hspace{1em} $\square$ Regression is for discrete labels, classification for real valued ones.  

\textbf{Question 3.} What is the difference between $L_1$ and $L_2$ regularization?

\hspace{1em}$\square$ $L_1$ yields sparser solutions.

\hspace{1em} $\square$ $L_2$ yields sparser solutions.

\hspace{1em} $\square$ none - they differ only in few practical applications.  

\textbf{Question 4.} In the bias-variance trade-off, what does high variance imply?

\hspace{1em} $\square$ The analyzed data shows high variance.

\hspace{1em} $\square$ The clf is overly complex and potentially overfits.

\hspace{1em} $\square$ The data is likely to be classified fair (e.g., low bias).  

\textbf{Question 5.} Why is Naive Bayes naive?

\hspace{1em} $\square$ Due to historic reasons.

\hspace{1em} $\square$ Due to the assumption that all features are independent.

\hspace{1em} $\square$ Because the application is simple and straight-forward. 

\textbf{Question 6.} What is cross-validation?

\hspace{1em} $\square$ Training on one task and then transferring the model to another task.

\hspace{1em} $\square$ Splitting the dataset and training/evaluating on different subsets.

\hspace{1em} $\square$ A method to reduce overfitting or choosing hyper-parameters.

\begin{table*}
\footnotesize
\caption{Final set of codes for the interviews.}\label{table:interviewcodes}
\centering
\begin{tabular}{  l | l | l | l  }
\toprule
	\textbf{A. AML attacks} & \textbf{D. security defenses} & \textbf{G. organization} & \textbf{L. perception} \\ 
	A.1 poisoning & D.1 sandboxing & G.1 ML role in project & L.1 security externalized\\
	A.2 evasion & D.2 access control & G.2 security role in project & L.2 AML feature not bug\\
	A.3 model stealing & D.3 development policy & G.3 other role on project & L.3 doubting attacker \\
	A.4 reverse engineering & D.4 server register & G.4 legal constraints & L.4 believing defense is effective \\ 
	A.5 membership inference & D.5 security testing & G.5 technical dept of ML & L.5 has not encountered threat \\
	A.6 availability & D.6 data anonymization & \textbf{H. customer} & L.6 attacks too specific \\ 
    \textbf{B. AML defenses} & D.7 input data format restrictions & H.1 requirements & L.7 insecurity about AML \\ 
	B.1 retraining & E. \textbf{pipeline elements} & H.2 privacy relevant data & L.8 unspecific attack \\ 
	B.2 interpretability & E.1 training & \textbf{I. cloud} & L.9 holistic attacker specificity \\ 
	B.3 basic models & E.2 design & I.1 used for security & L.10 pipeline specific defense \\ 
    B.4 ensemble & E.3 model & I.2 used but potential security risk & L.11 importance of data \\ 
	B.5 human in the loop & E.4 data & I.3 not used because of security & L.12 high level perspective \\ 
	B.6 regularization & E.5 data labelling & I.4 neutral & L.13 coding perspective \\ 
	B.7 own implementation & E.6 data collection & \textbf{J. relevance} & \\ 
	B.8 on purpose & E.7 data preprocessing & J.1 mentioning AML & \\ 
	\textbf{C. security threats} & E.8 feature extraction & J.2 security low priority & \\ 
	C.1 data capturing & E.9 testing & J.3 AML low priority & \\ 
	C.2 access & E.10 deployment & J.4 encountered security issue & \\ 
	C.3 data breach & E.11 API & \textbf{K. confusion} & \\ 
	C.4 code breach & E.12 database & K.1 across ML attacks & \\ 
	C.5 libraries & \textbf{F. pipeline properties} & K.2 security and AML & \\ 
	C.6 denial of service & F.1 iterative & K.3 vagueness of concepts & \\ 
	C.7 SDK & F.2 several within project & K.4 what security means & \\ 
	C.8 customer & & & \\ \bottomrule
\end{tabular}
%} 
\vspace{-1em}
\end{table*}

\textbf{Question 7.} What are kernels in machine learning?

\hspace{1em} $\square$ Essentially similarity functions. 

\hspace{1em} $\square$ A part of SVM, potentially yielding non-linear SVM.

\hspace{1em} $\square$ A specific instance of a similarity function used in SVM.  

\textbf{Question 8.} What is pruning? 

\hspace{1em} $\square$ Deletion of for example weights in a model.

\hspace{1em} $\square$ Deletion of specific points of the data.

\hspace{1em} $\square$ A technique to get a smaller from a large model \\ with similar performance.  

\vspace{0.5em}
To conclude the study, we will ask you to rate your background knowledge on attacks \emph{before} this study according to the following four classes:

\hspace{1em} \textbf{Familiar.} Your are familiar with this concept, and can write down the mathematical formulation.

\hspace{1em} \textbf{Dabbled in.} You could explain in a five minute talk what the concept is about.

\hspace{1em} \textbf{Heard of.} You have heard of the concept and you could put it into context if necessary.

\hspace{1em} \textbf{Never heard.} You did not know about this concept before this survey.

 For each concept, please tick \textbf{one} box. \emph{The original questionnaire was formatted as table. To ease readability, we list them as questions here.} 
 
 \textbf{Evasion / adversarial examples}. \\
$\square$ familiar \hspace{1em} $\square$ dabbled in \hspace{1em}  $\square$ heard of \hspace{1em}  $\square$ never heard

 \textbf{Poisoning / backdooring} \\
$\square$ familiar \hspace{1em} $\square$ dabbled in \hspace{1em}  $\square$ heard of \hspace{1em}  $\square$ never heard

 \textbf{Model stealing } \\
$\square$ familiar \hspace{1em} $\square$ dabbled in \hspace{1em}  $\square$ heard of \hspace{1em}  $\square$ never heard

    \textbf{Model reverse engineering }  \\
$\square$ familiar \hspace{1em} $\square$ dabbled in \hspace{1em}  $\square$ heard of \hspace{1em}  $\square$ never heard
   
    \textbf{Neural trojans } \\
$\square$ familiar \hspace{1em} $\square$ dabbled in \hspace{1em}  $\square$ heard of \hspace{1em}  $\square$ never heard

    \textbf{Adversarial initialization}  \\
$\square$ familiar \hspace{1em} $\square$ dabbled in \hspace{1em}  $\square$ heard of \hspace{1em}  $\square$ never heard

\section{Final set of codes}\label{app:codes}
The final set of codes for the interviews is depicted in Table~\ref{table:interviewcodes}, the codes for the drawings in Table~\ref{table:drawingcodes}.

\section{Structural and functional components} \label{sec:strucfunc}
We found structural and functional components in our participants' mental models. Structural components cover multiple, constituting entities that an individual perceives as relevant within a given application. In interaction with an ML system, functional components describe an individual's perception of the relations between the structural elements. As intended, the structure of our interview and drawing task (Appendix~\ref{app:interview}) allowed to investigate these properties on the level of the ML pipeline, of the attack vectors as well as of the defenses. 

\subsection{ML pipeline} All participants distinguish clearly separable elements within their ML workflow. The specific composition of these steps defines the structure of a certain ML pipeline. For two participants, this structure reflects the ML pipeline that we introduced in Figure~\ref{fig:pipeline}. When asked to sketch the kind of pipeline applied, \Sub{4} talked about \quoteSub{data}, \quoteSub{training}, \quoteSub{testing}, and \quoteSub{visualization}. We argue that these structural components serve as a scaffold for an individual's mental model. Interestingly, the mental models of 12 out of 15 participants covered additional components that we did not expect prior to the study. The sketches of \Sub{3}, \Sub{7}, and \Sub{11} (Figure~\ref{fig:SecLEvel11}), for example, contain explicit elements for data capturing. \Sub{1}, 
\Sub{9}, \Sub{12}, as well as \Sub{20} included dedicated elements representing a specific database to their drawing. Five participants also highlighted structural elements within the deployment environment during the interviews. \Sub{14}, for example, specified on an API for deployment \quoteSub{on several kinds of hardware architectures}. Analogously, \Sub{1} described an API that \quoteSub{can be used to allow the user to interact with the models} 
Hence, these structural elements concerning data and deployment seem to be of importance for the corresponding mental models. However, the perception of industrial practitioners does no only focus on these structural components but also covers functional aspects. \Sub{6} for instance stated that his ML pipeline \quoteSub{forks into a number of different directions and there are also interactions between the different components}. In the corresponding sketch, multiple arrows within and across specific ML models indicate this interconnection of single components. Other drawings include this functional perspective through straight lines connecting the structural components, 
arrows connecting some of the structural components in a subsequent manner (e.g. \Sub{14}), and arrows connecting all structural components in a subsequent manner (\Sub{18} in Figure~\ref{fig:SecLEvel18}).

\begin{table*}[t]
\caption{Final set of codes for the drawings.}\label{table:drawingcodes}
\footnotesize
\centering
\begin{tabular}{  l | l | l | l | l }
\toprule
	\textbf{A. pipeline elements} & \textbf{B. pipeline properties} & \textbf{C. named explicitly} & \textbf{D. attacks} & \textbf{E. drawing} \\ 
	A.1 training & B.1 iterative & C.1 hardware & D.1 no attacks & E.1 boxes\\ 
	A.2 design & B.2 linear & C.2 software & D.2 poisoning & E.2 symbols\\ 
	A.3 model & B.3 abstracted & C.3 human & D.3 evasion & E.3 inner/outer \\ 
	A.4 data & B.4 several & C.4 privacy sanititzation & D.4 membership inference & E.4 flow within pipeline \\ 
	A.5 data labelling & B.5 explainable & C.5 output & D.5 libraries & E.5 workflow embedding \\ 
	A.6 data collection & B.6 MLaaS & C.6 classification & D.6 data collection & E.6 attacks graphical \\ 
    A.7 data preprocessing &  & C.7 server & D.7 input/output & E.7 attacks words \\ 
	A.8 feature extraction &  & & D.8 unspecific attack & E.8 attacks causal \\ 
	A.9 testing & & & D.9 defenses & E.9 attacks pointwise \\ 
	A.10 deployment & 	&& D.10 exit points & \\ 
    A.11 deployment environment	& && D.11 input points & \\  \bottomrule
\end{tabular}
\end{table*}

\begin{figure*}[ht]
\centering
    \includegraphics[trim=0 0.5 0 4, clip, width=1.9\columnwidth]{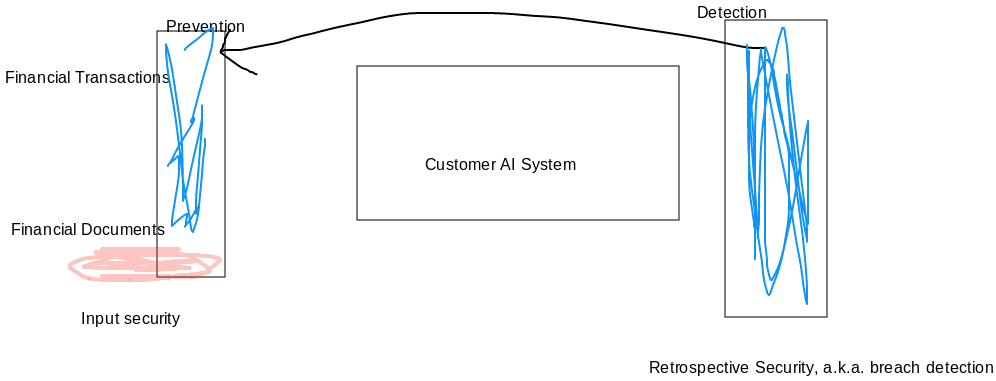}
    \caption{Drawing of \Sub{10}. Important components of the workflow added in blue, possible 
           starting points for attacks in red.}\label{fig:SpecAttacks10}
\end{figure*}

\subsubsection{Attack vectors} The identified structural and functional components seem to be similarly relevant for mental models on attack vectors. For any kind of ML-specific threat, participants were able to precisely locate where they situated the corresponding, structural starting point. These have been specifically named during the interview and sketched via labelled arrows (e.g. Figure~\ref{fig:SecLEvel11}, \Sub{11}), additional annotations (\Sub{11}, \Sub{15}), highlighted parts of potentially vulnerable pipeline components (e.g. Figure~\ref{fig:SpecAttacks10}, \Sub{10}) or as entire steps within a given ML workflow that have been marked as vulnerable (\Sub{9}, \Sub{20}). Strikingly, we saw a wide overlap in the perception of potential focal starting points for attack vectors. Study participants considered the model itself, the input of their ML pipeline, or the deployment environment to be particularly vulnerable. Figure~\ref{fig:SpecAttacks16} (\Sub{16}) shows this for the latter. When confronted with poisoning and reverse engineering attacks, \Sub{16} marked the input and output of his pipeline as possible starting points for threats (purple rectangles) and talked about how a competitor could \quoteSub{screw our labeled dataset} or a customer might \quoteSub{ask a lot of questions to the API}. However, the perception of attack vectors did also cover functional components. \Sub{1}, for example, depicted the causal sequence of a \quoteSub{data injection attack} as three consecutive red arrows connecting different components of his ML pipeline. 
This is all the more relevant, as \Sub{1} provided such a functional explanation and drawing for each of the attack vectors we presented to him. His mental models, hence, clearly seem to contain functional components. This is also the case for \Sub{16}, who similarly provided explanations on the functional evolvement of certain attacks within his workflow and even added corresponding functional elements to his sketch (blue and red arrows in Figure~\ref{fig:SpecAttacks16}).

\subsubsection{Defenses} Although we found participants' defenses explanations and sketches to be rather sparse, structural and functional properties are also relevant for the corresponding mental models. As visible in the sketch of \Sub{18}, defenses are often thought of as structurally bound to specific components of a workflow/pipeline (Figure~\ref{fig:SecLEvel18}, \Sub{18}). Data (\Sub{14}), training (\Sub{6}) and the models themselves (\Sub{10}) have been specifically named as focal points for implementing defenses. In the case of defenses implemented at the model component, \Sub{14} stated to \quoteSub{regularize in a way that makes it less sensitive to an adversary}.
Hence, these implemented defenses are cognitively attached to the classifier as a focal pipeline component. However, security mental models also contain functional properties. In the case of human-in-the-loop-defenses, for example, \Sub{14} stated to send certain classifications \quoteSub{back to the data sets to train a second version of the algorithm} if the output confidence for certain data exhibited high entropy. 
This is depicted in the corresponding sketch by an arrow pointing from a rectangle with the caption \quoteSub{CPU} at the end of the pipeline to \quoteSub{raw data} (initial step of the pipeline). Similarly, \Sub{7}, a participant working in video surveillance, explained the defense they had implemented to secure the transfer of input data (from cameras and on-site computers) into their pipeline: \quoteSub{This can only go out, never go in. [...] Nothing from the internet can connect to that server}. Industrial practitioners, hence, perceive defenses as containing functional components to unfold their full effect. 

\subsubsection{Summary} We conclude that mental models in AML contain of structural components which are cognitively put into (internal) relation. However, the specific unfolding of these internal conceptual representations seems to depend on the corresponding application and its underlying ML pipeline.